\documentclass[a4paper,twoside]{article}
\newcommand{\affil}[1]{$^{\rm #1}$}
\usepackage[authoryear]{natbib}
\bibpunct{(}{)}{;}{a}{}{,}
\usepackage{graphicx}
\usepackage{rotating}
\usepackage{amsmath}
\usepackage{amssymb}
\usepackage{mathabx}
\date{} 
\title{\large\bf\flushleft On the Estimation of Confidence Intervals for Binomial Population Proportions in Astronomy: The Simplicity and Superiority of the Bayesian Approach}
\author{\parbox{\textwidth}{\flushleft
\vspace{-0.5cm}
{\it Ewan Cameron\affil{A,B}}\\
\vspace{0.4cm}
{\small \affil{A}\,Department of Physics, Swiss Federal Institute of Technology (ETH Zurich), CH-8093 Zurich, Switzerland}\\
{\small \affil{B}\,Email: cameron@phys.ethz.ch}}}
\begin{document}
\twocolumn[
\begin{changemargin}{.8cm}{.5cm}
\begin{minipage}{.9\textwidth}
\vspace{-1cm}
\maketitle
%
%
\small{\bf Abstract:} I present a critical review of techniques for estimating confidence intervals on binomial population proportions inferred from success counts in small-to-intermediate samples.  Population proportions arise frequently as quantities of interest in astronomical research; for instance, in studies aiming to constrain the bar fraction, AGN fraction, SMBH fraction, merger fraction, or red sequence fraction from counts of galaxies exhibiting distinct morphological features or stellar populations.  However, two of the most widely-used techniques for estimating binomial confidence intervals---the `normal approximation' and the Clopper \& Pearson approach---are liable to misrepresent the degree of statistical uncertainty present under sampling conditions routinely encountered in astronomical surveys, leading to an ineffective use of the experimental data (and, worse, an inefficient use of the resources expended in obtaining that data).  Hence, I provide here an overview of the fundamentals of binomial statistics with two principal aims: \textsc{(i)} to reveal the ease with which (Bayesian) binomial confidence intervals with more satisfactory behaviour may be estimated from the quantiles of the beta distribution using modern mathematical software packages (e.g.\ \textsc{R}, \textsc{matlab}, \textsc{mathematica}, \textsc{IDL}, \textsc{python}); and \textsc{(ii)}  to demonstrate convincingly the major flaws of both the `normal approximation' and the Clopper \& Pearson approach for error estimation.


\medskip
\medskip
\end{minipage}
\end{changemargin}
]
\small

\section{Introduction}
One problem frequently encountered in astronomical research is that of estimating a confidence interval (CI) on the value of an unknown population proportion based on the observed number of success counts in a given sample.  The unknown population proportion may be, for instance, the intrinsic fraction of barred disk galaxies at a specific epoch to be inferred from the observed number of barred disks in a volume-limited sample (e.g.\ \citealt{elm90,van02,cam10,nai10}), with the corresponding binomial CI used to evaluate the hypothesis that the bar fraction changes with redshift relative to a local benchmark (e.g.\ \citealt{cam10}). Experiments to investigate the role of mass and environment in quenching star-formation via measurement of the galaxy red sequence fraction (e.g.\ \citealt{bal06,hes10,ilb10}), or to investigate whether or not major mergers were more frequent at high redshift via measurement of the close-pair/asymmetric fraction (e.g.\ \citealt{dep05,con08,lop10}), also routinely present this class of problem.

However, the two most commonly used methods for estimating CIs on binomial population proportions---the `normal approximation' and the \citet{clo34} approach---exhibit significant flaws under routine sampling conditions (cf.\ \citealt{vol93,san98,bro01,bro02}).  In particular, the `normal approximation' (also called the `Poisson error') may systematically under-estimate the CI width necessary to provide coverage at the desired level, especially for small samples, but even for rather large samples when the true population proportion is either very low or very high.  If used na\"ively the `normal approximation' has the potential to mislead one into over-stating the significance of one's inferences concerning the physical system under study formulated on the basis of the observed data.

Astronomers aware of these flaws in the `normal approximation' often adopt the alternative \citet{clo34} approach to CI estimation by way of reference to the CI tables in \citet{geh86}.  Unfortunately, the \citet{clo34} approach suffers from the opposite problem to that of the `normal approximation'---namely, a systematic over-estimation of the CI width required to provide the desired coverage \citep{clo34,ney35,geh86,agr98}.  In scientific research this over-estimation of the statistical measurement uncertainties may mislead one into placing insufficient confidence in the experimental outcomes, resulting in an inefficient use of the measured data (and, hence, the resources expended in obtaining that data).  Indeed, it has been well argued by \citet{agr98} that in many practical applications even the `normal approximation', despite its flaws, is preferable to the \citet{clo34} approach.

Fortunately, there exist a multitude of alternative methods for generating CIs on binomial population proportions, many of which exhibit far more satisfactory behaviour than either the `normal approximation' or the \citet{clo34} approach---see \citet{agr98} and \citet{bro01} for various examples.  Here I review both the theory and application of one of these methods---use of the beta distribution quantiles---deriving from a simple Bayesian analysis in which a uniform (`non-informative') prior is adopted for the true population proportion (e.g.\ \citealt{gel03}).  As I will demonstrate, the beta distribution generator for binomial CIs is both theoretically well-motivated and easily applied in practice using widely available mathematical software packages (e.g.\ \textsc{R}, \textsc{matlab}, \textsc{mathematica}, \textsc{IDL}, \textsc{python}).  Ultimately, I advocate strongly that this strategy for estimating binomial CIs be adopted in future studies aiming to constrain fundamental population proportions in astronomical research (e.g.\ the galaxy bar fraction, red sequence fraction, or merger fraction)---especially for samples intrinsically of small-to-intermediate size, or when the subdivision of larger samples for analytical purposes produces sparsely populated data bins.

\section{The Binomial Distribution}
In probability theory any experiment for which there are only two possible random outcomes---\textit{success}, occurring with probability, $p$, or \textit{failure}, occurring with probability, $q=(1-p)$---is referred to as a \textit{Bernoulli trial}.  Examples of Bernoulli trials in astronomical research may include asking whether or not a randomly sampled galaxy is barred, red-sequence, or merging.  The probability, $P$, of observing a particular number of successes, $k$, in a series of $n$ independent Bernoulli trials (with common success probability, $p$) is governed by the \textit{binomial} probability function\footnote{One may note that the correct terminology in a statistical context is actually `binomial probability \textit{mass} function', owing to the discrete nature of the binomial distribution, i.e., that there exist a finite number of possible $k$ values (the integers from 0 to $n$, inclusive) to which non-zero probabilities may be assigned.  (As distinct from the alternative case of a `probability \textit{density} function', such as the Bayesian posterior probability distribution for $p$ considered in Section \ref{section3}, for which non-zero probabilities may only be assigned to measurable intervals on the real number line, and not individual---or even countable sets of---real numbers.)  Nevertheless, to avoid any confusion with the more commonly-used definition of the term, `mass function', in astronomy I adopt the shorter expression, `binomial probability function', herein.}:
\begin{equation}\label{binom}
P(k, n, p) = \binom{n}{k} p^k q^{n-k}
\end{equation}
where $0 \le k \le n$, $k\in\mathbb{Z}$ (an integer), and 
\[
\binom{n}{k} = \frac{n!}{k!(n-k)!}
\]
(see, for example, \citealt{qui78}).  Note that the probabilities given by the $n+1$ possible values of $k$ correspond to the $n+1$ terms of the binomial expansion of $(p+q)^n$.  The number of barred systems counted in a given sample of disk galaxies is a classic example of a binomially-distributed variable in astronomy.  The corresponding expectation value for the number of successes is $\sum_{k=0}^{n} k \times P(k,n,p) = np$ with a variance of $\sum_{k=0}^{n} k^2 \times P(k,n,p) = npq$.  Moreover, the expectation value for the \textit{fraction}, $k/n$, of successes is equal to the Bernoulli trial success probability (also referred to as the `underlying population proportion'), $p$, and its variance is $pq/n$.

\paragraph{An intermission: Just what is a confidence interval?}
As explained eloquently by both \citet{kra91} and \citet{ros03}, there is a fundamental difference between the `classical' and `Bayesian' definitions of the term `\textit{confidence interval}'.  In classical statistical theory a binomial CI is defined as a pair of random variables, $P_l$ and $P_u$, (with each random variable necessarily a finite, real-valued, measurable function; cf.\ \citealt{rao06}) operating on the set of all possible experimental outcomes, $\theta = \{k:0 \le k \le n, k\in\mathbb{Z}\}$, such that if the experiment were to be repeated by a sufficiently large number of independent observers then the \textit{fraction of observers} for whom the true value of the underlying population proportion is covered by their realisation of these random variables---i.e., for whom $P_l (\theta_i) =p_l < p < p_u=P_u(\theta_i)$---is guaranteed to converge to (at least) a specific value, $c$, termed `the confidence level'.  In the Bayesian paradigm, on the other hand, the underlying population proportion is treated as an unknown model parameter and the binomial CI defined as an interval, $(p_l, p_u)$, to which the experimenter believes may be assigned a \textit{probability}, $c$, of containing the true value of $p$, based upon consideration of the likelihood function for $p$ given the experimental data and the strength of any \textit{a priori} beliefs/expectations regarding the system under study.  (Indeed, acknowledging the significant conceptual differences between these alternative approachs to the binomial CI, the term `credible interval' is often used instead in Bayesian analysis to avoid confusion with the classical nomenclature.)  Importantly, as noted by \citet{kra91}, regardless of one's philosophical position regarding these two statistical systems, ``the Bayesian definition of confidence intervals reflects common astronomical usage better than the classical definition''.

\begin{figure*}
\begin{center}
\includegraphics[width=7.5cm]{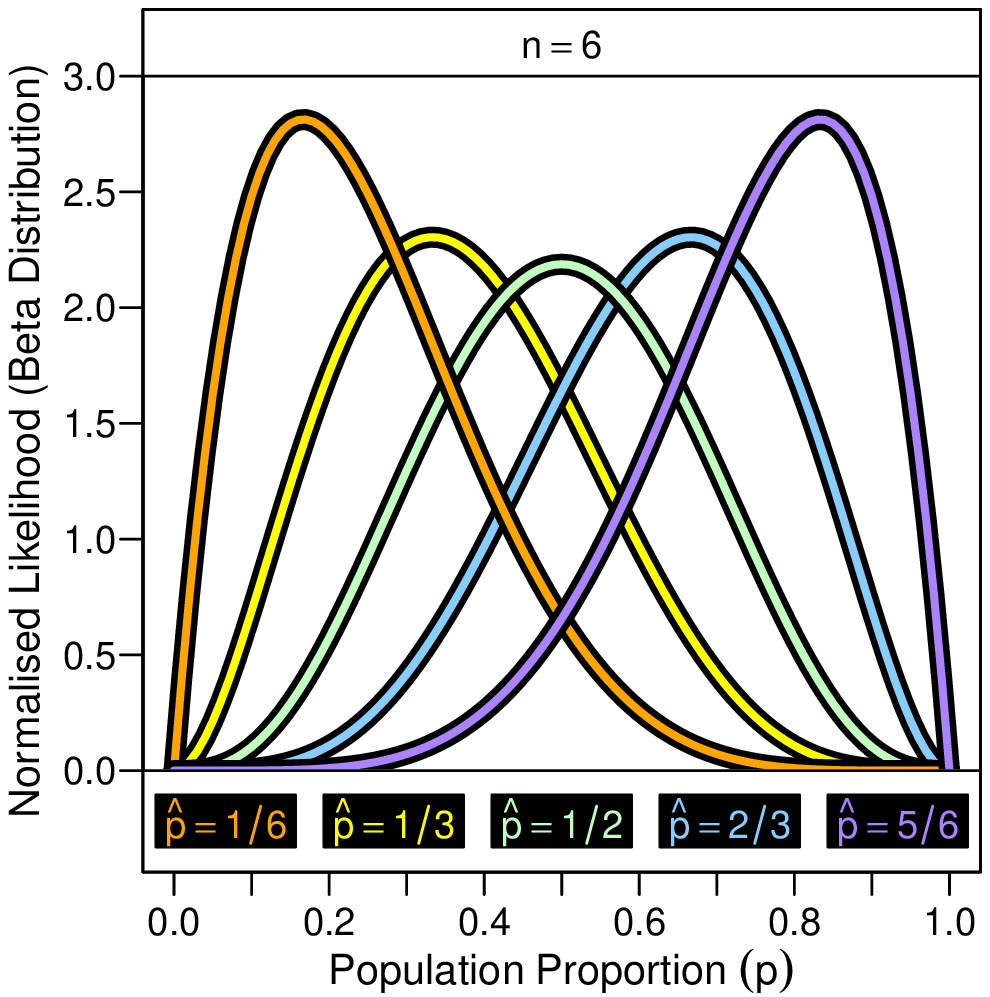} \includegraphics[width=7.5cm]{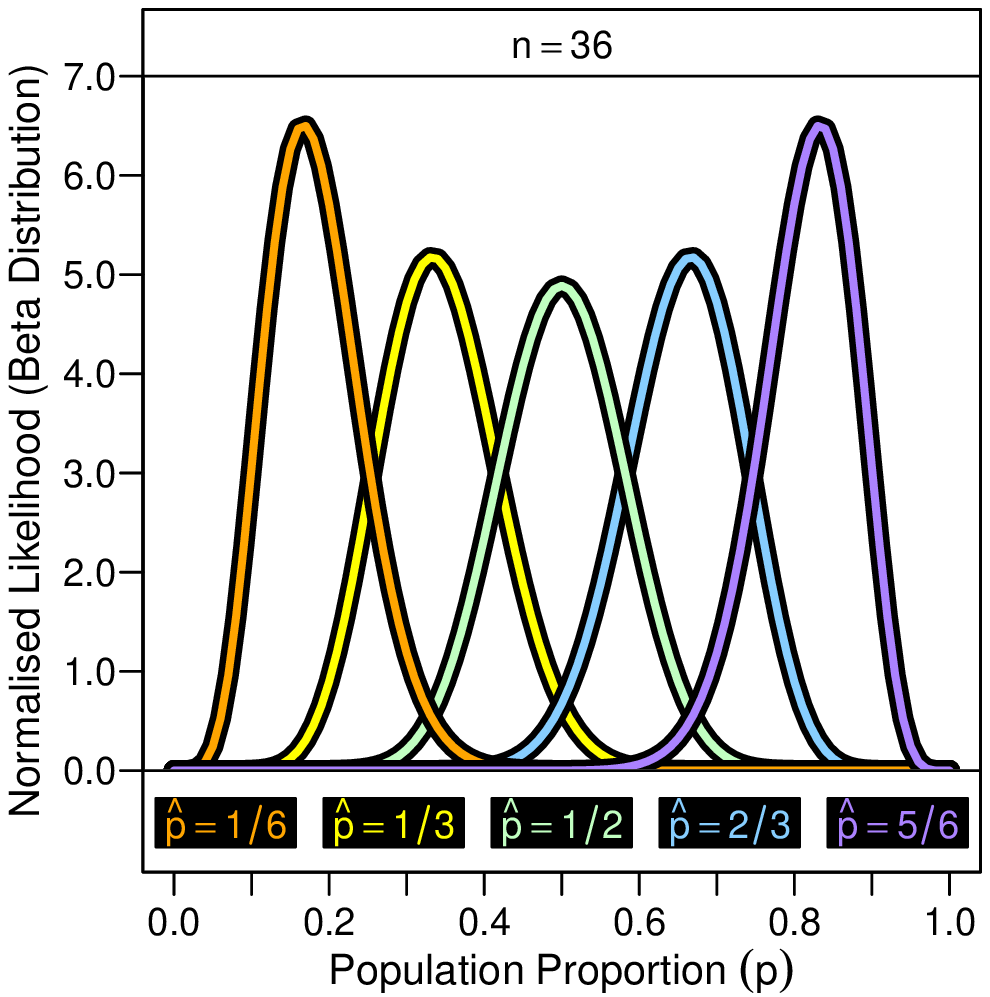}
\caption{Example likelihood functions for the true value of the underlying population proportion, $p$, given five `measured' success fractions, $\hat{p}=k/n$, for samples of sizes $n=6$ (left panel) and $n=36$ (right panel).  In each case the shape of the curve is given by the beta distribution with shape parameters as specified by Equation \ref{betaeqn}.  The asymmetric nature of this likelihood function in the small sample size regime is clearly evident amongst the $n=6$ examples, as is its convergence in the intermediate-to-large sample size regime towards a narrower, more symmetric, (pseudo-)normal distribution amongst the $n=36$ examples.\label{fig1}}
\end{center}
\end{figure*}

Of the three binomial CI generators discussed in this review only that attributed to \citet{clo34} is consistent with the classical definition for all possible values of the underlying population proportion and sample size.  However, I will argue that (at least) in the case of the binomial distribution Bayesian CIs provide generally more satisfactory behaviour for astronomical purposes than their classical counterparts, even when evaluated against a performance diagnostic based on the classical definition---namely, the coverage fraction (or `effective coverage') at given $p$ and $n$.

\section{The Beta Distribution Generator for Binomial Confidence Intervals}\label{section3}
In astronomical data analysis it is standard practice to adopt the measured success fraction (also referred to as the `observed population proportion'), $\hat{p} = k/n$, as one's `best guess' of the underlying population proportion.  In statistical terms, $\hat{p}$ is employed as a \textit{point estimator} for $p$.  The likelihood of observing the result, $\hat{p} = k/n$, for a given value of $p$ is, of course, proportional to $p^k q^{n-k}$.  Normalisation of this likelihood function over $0 < p < 1$ defines a `beta distribution' with integer parameters, $a = k+1$ and $b = n-k+1$:
\begin{equation}\label{betaeqn}
B(a,b) = \frac{(a+b-1)!}{(a-1)!(b-1)!}p^{a-1}q^{b-1}
\end{equation}
where $q=1-p$ (e.g.\ \citealt{gel03,ros03}).  Differentiation of this likelihood function reveals that our best guess, $\hat{p}$, is in fact the maximum likelihood estimator of $p$.\footnote{Technically, when $\hat{p}=0$ (or 1) the likelihood function for $p$ has no zero first derivative on the open interval, $(0,1)$, although the function itself is indeed strictly increasing as $p\rightarrow 0$ (or 1).  In this case one may choose to adopt the median (50\% quantile) of the (beta distribution) likelihood function as one's best guess for $p$, or else to compute a `one-sided' confidence interval bounding $p$ instead.  In either case, one proceeds along similar principles.}  The characteristic shape of the (beta distribution) likelihood function for $p$ is illustrated in Figure \ref{fig1} at a variety of `measured' success fractions for samples of sizes $n=6$ (left panel) and $n=36$ (right panel).  At small $n$, the likelihood function for $p$ is markedly asymmetric (except where $\hat{p} = 1/2$), but at intermediate $n$ it is visibly converging towards a narrow, symmetric, (pseudo-)normal distribution---the motivation behind the `normal approximation' discussed in Section \ref{section4}.

Given no \textit{a priori} knowledge to inform one's expectations regarding the experimental outcome one may suppose that all values of $p$ are equally `probable'.  Formally, this condition is characterised via the Bayes-Laplace uniform prior, for which $P_\mathrm{prior}(p)=1$ over $0 < p < 1$.  Application of Bayes' theorem under this assumption allows one to treat the normalised likelihood function for $p$ as a posterior probability distribution.  Thus, the quantiles of the beta distribution from Equation \ref{betaeqn} may be used directly to estimate (Bayesian) confidence intervals on the underlying population proportion given the observed data.\footnote{Astronomers familiar with the work of \citet{bur03} on binarity in brown dwarfs may be familiar with the algorithm for recovering confidence intervals on $p$ given in their Appendix, which is in fact equivalent to the Bayesian approach with uniform prior presented here (although \citealt{bur03} make no explicit reference to either Bayes' theorem or the beta distribution).}  Specifically, the lower and upper bounds, $p_l$ and $p_u$, defining an `equal-tailed' (or `central') interval for $p$ at a nominal confidence level of $c = 1-\alpha$ are given by the quantiles:
\begin{equation}\label{betad}
\int^{p_l}_0 B(a,b) dp = \alpha/2 \mathrm{\ and\ } \int_{p_u}^1 B(a,b) dp = \alpha/2 \mathrm{\ .}
\end{equation}
Note that the bounds of this `equal-tailed' interval (which partition the probability of $p$ greater than $p_u$ equal to that of $p$ less than $p_l$) will be necessarily asymmetric about the maximum likelihood value, $\hat{p}$, (except at $\hat{p}=1/2$) owing to the asymmetric nature of the (beta distribution) likelihood function for $p$ (shown in Figure \ref{fig1}).  As I will demonstrate below, binomial CIs generated in this manner have one rather desirable property, not shared by either the `normal approximation' or the \citet{clo34} approach---namely, their \textit{mean} effective coverage is consistently very close to the nominal confidence level, even in the small sample size regime.

\begin{figure*}
\begin{center}
\includegraphics[width=15cm]{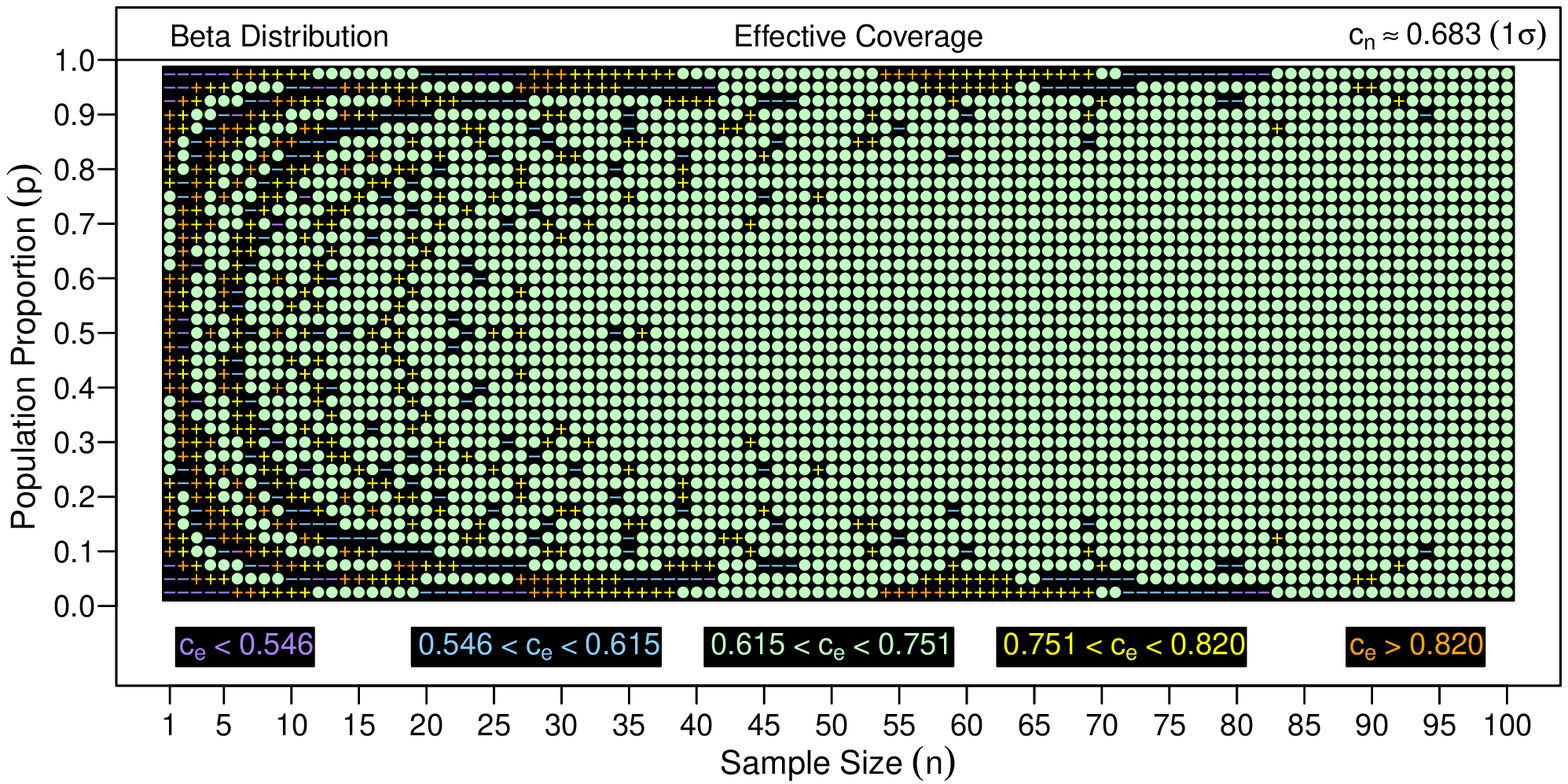} \includegraphics[width=15cm]{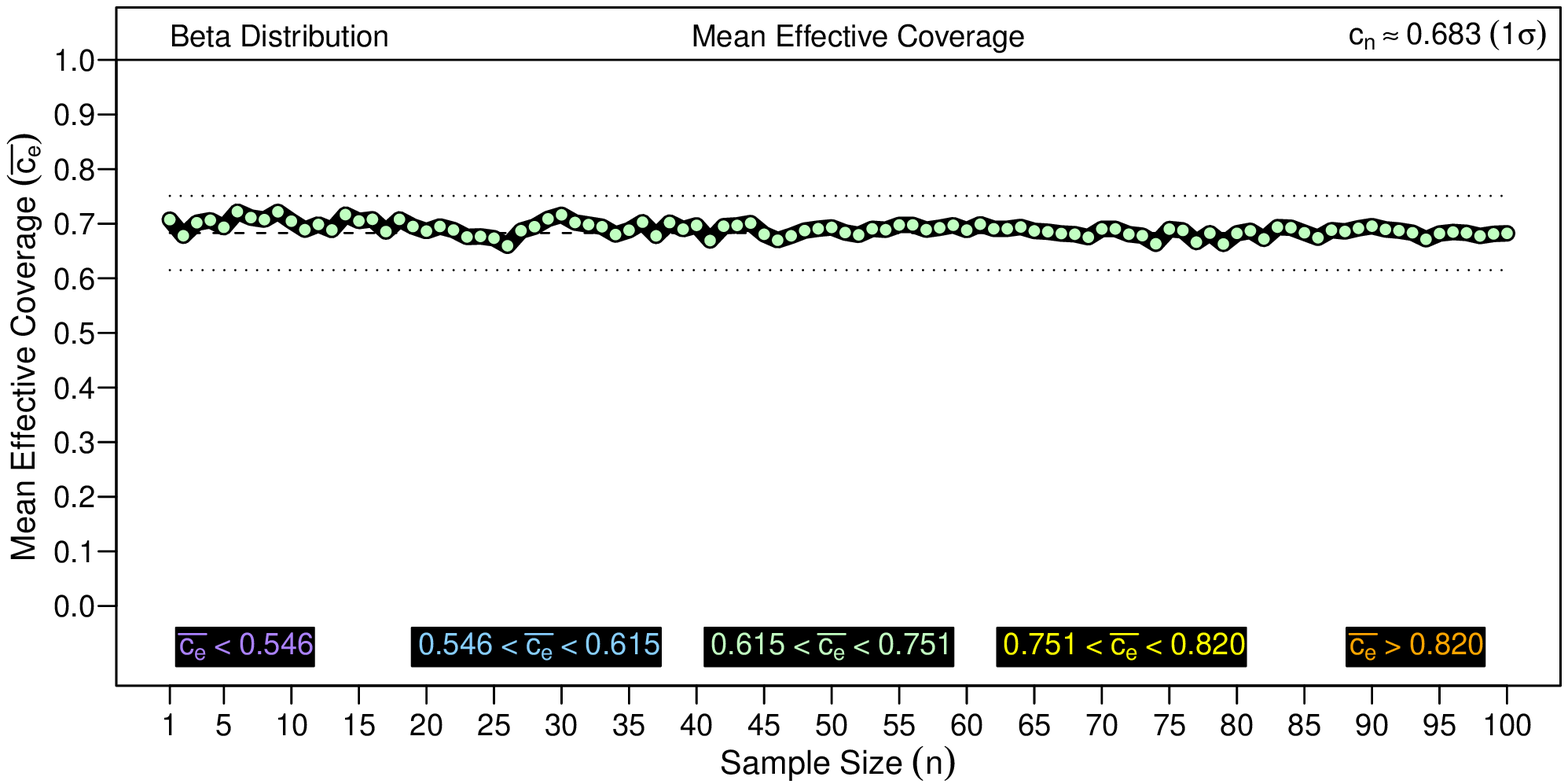}
\caption{The effective coverage, $c_e$, of confidence intervals on the binomial population proportion generated from quantiles of the beta distribution at a nominal level of $c_n \approx 0.683$ (1$\sigma$) over the range $0.025 \le p \le 0.975$ and $1 \le n \le 100$ (upper panel).  Averaging the measured $c_e$ values uniformly over all $p$ at each $n$ returns the \textit{mean} effective coverage as a function of sample size (lower panel).\label{fig2}}
\end{center}
\end{figure*}

In the upper panel of Figure \ref{fig2} I examine first the effective coverage, $c_e$, of `equal-tailed' binomial CIs defined via the beta distribution for a range of population proportions and sample sizes ($0.025 \le p \le 0.975$ and $1 \le n \le 100$) at a nominal level of $c_n \approx 0.683$ (1$\sigma$)---with the effective coverage defined as the fraction of samples drawn from the binomial probability function with given $p$ and $n$ for which the corresponding realisation of the CI under investigation encompasses the true population proportion.  Thus, the effective coverage fractions, $c_e$, presented here are computed as the sum of all binomial probabilities, $P(k,n,p)$ over $\{k:0 \le k \le n,k\in\mathbb{Z}\}$, for which the triad, $\{k$,$n$,$p\}$, produces a confidence interval, $(p_l,p_u)$, containing (covering) $p$. 

One of the most striking features of this plot is the remarkable sensitivity of the effective coverage to the true underlying population proportion and sample size.  This so-called `oscillation signature' is an inherent property of \textit{all} deterministic (i.e., non-randomising) generators for binomial CIs, arising from the discreteness of the binomial distribution.\footnote{\citet{bro01} describe the `oscillation signature' as the challenge of `lucky $p$, lucky $n$'---namely that for certain (`lucky') combinations of underlying population proportion and sample size there exist two almost equally likely $\hat{p}$ values closely straddling the true $p$.  For instance, if $p=1/5$ and one has a sample of size $n=3$ the possible $\hat{p}$ values are 0, 1/3, 2/3, and 1, occurring with frequencies 0.512, 0.384, 0.096, and 0.008, respectively.  Tailoring a binomial CI specifically to this situation, one could define $p_l=\hat{p}-2/15-\epsilon$ and $p_u=\hat{p}+1/5+\epsilon$ (with $\epsilon$ an arbitrarily small constant necessary to ensure $p$ is contained within the open interval, $(p_l,p_u)$, for $k=0$ and 1), returning an effective coverage of $c_e = 0.512+0.384=0.896$.  However, applying the same CI generator to a system with $p=1/3$ (and again $n=3$) for which the possible $\hat{p}$ values occur with frequencies 0.296, 0.444, 0.222, and 0.037 (rounded to 3 decimal places), one obtains an effective coverage of only $c_e=0.444$!  For further discussion of the impact of the `oscillation signature' on binomial CIs the interested reader is referred to \citet{agr98} and \citet{bro01,bro02}.}  Despite these oscillations it is clear that the beta distribution CIs do achieve an effective coverage close to (or slightly greater than) the desired confidence level over the vast majority of the parameter space explored here.   Indeed, even at the extremes of $p \lesssim 1/6$ and $p \gtrsim 5/6$, where the oscillations are initially rather large, there is evidently a rapid increase in coverage stability with increasing sample size, such that the `oscillation signature' is vastly suppressed by $n \gtrsim 40$, and effectively eliminated (at least for $0.025 \le p \le 0.975$) by $n \gtrsim 80$ (unlike in the case of the `normal approximation' examined in Section \ref{section4}).

In the lower panel of Figure \ref{fig2} I examine the corresponding \textit{mean} effective coverage (averaged uniformly over $0.025 \le p \le 0.975$) as a function of sample size.  Whereas the effective coverage at given $p$ and $n$ shown in the upper panel is consistent with the classical notion of confidence interval performance, the \textit{mean} effective coverage may be considered a `Bayesian' CI performance diagnostic---i.e., if one really does hold all $p$ values equally `probable' \textit{a priori} then one's favoured CI generator should be at least expected to provide coverage consistent with the nominal level in the longterm average of all equivalent experiments.  Inspection of the lower panel of Figure \ref{fig2} confirms a very close agreement between the \textit{mean} effective coverage of the beta distribution CI generator and the nominal confidence level, independent of $n$. 

Most modern mathematical software packages provide robust, easy-to-use library functions for computing beta distribution quantiles  (e.g.\ the \textsc{qbeta} routine in \textsc{R}; the \textsc{Quantile} and \textsc{BetaDistribution} commands in \textsc{mathematica}; the \textsc{betaincinv} function in \textsc{matlab}; the \textsc{ibeta} function in \textsc{IDL}; or the \textsc{dist.beta.ppf} function in \textsc{python}).  Explicit code fragments demonstrating the implementation of these commands are provided in the Appendix to this paper, and I advocate strongly the use of these recipes for the computation of confidence intervals on binomial population proportions in future astronomical studies.  In Tables \ref{tableA1} and \ref{tableA2} in the Appendix I present compilations of `equal-tailed' CIs generated in this manner at nominal confidence levels of 1$\sigma$ and 3$\sigma$, respectively, for all possible observed success counts in sample sizes up to $n=20$.  These tables are intended both as a convenient reference for use directly in studies involving samples of 20 objects or less, and as a benchmark against which to confirm the correct implementation of the beta distribution CI generator for users newly adopting this technique.

\begin{figure*}
\includegraphics[width=7.5cm]{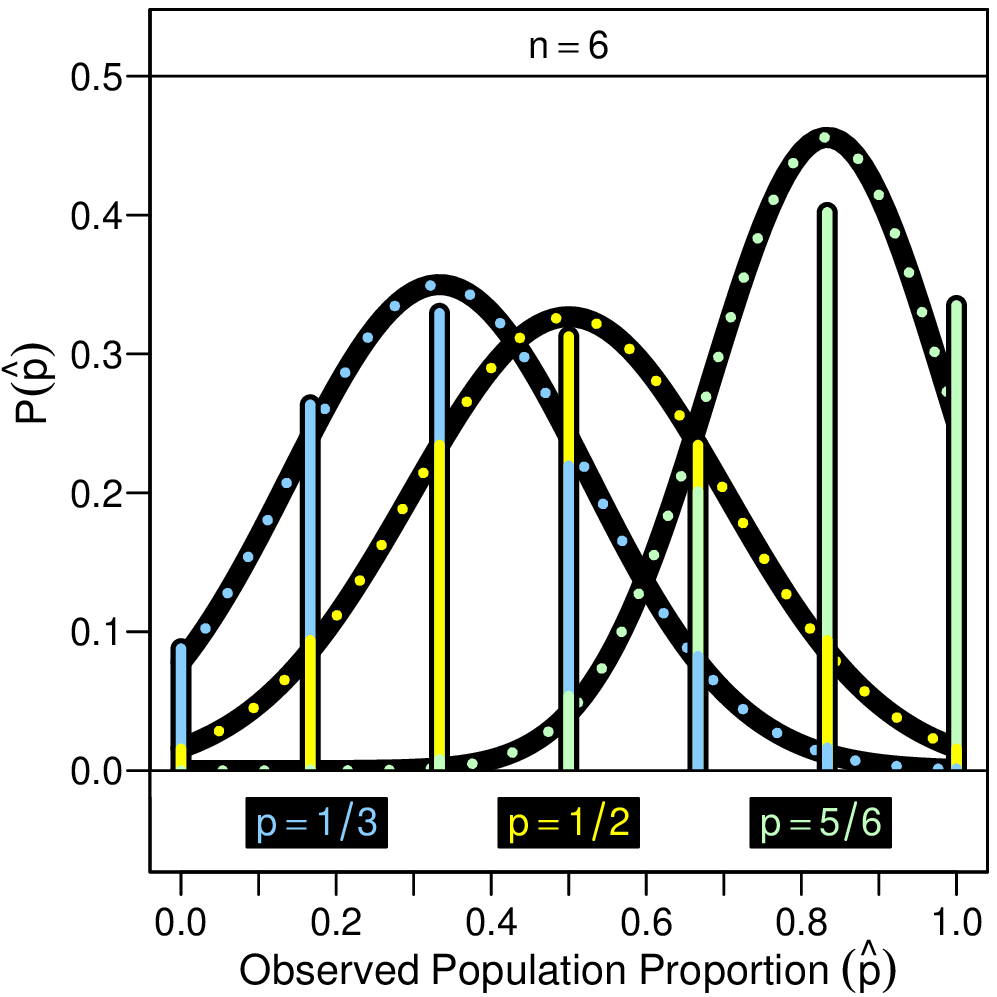} \includegraphics[width=7.5cm]{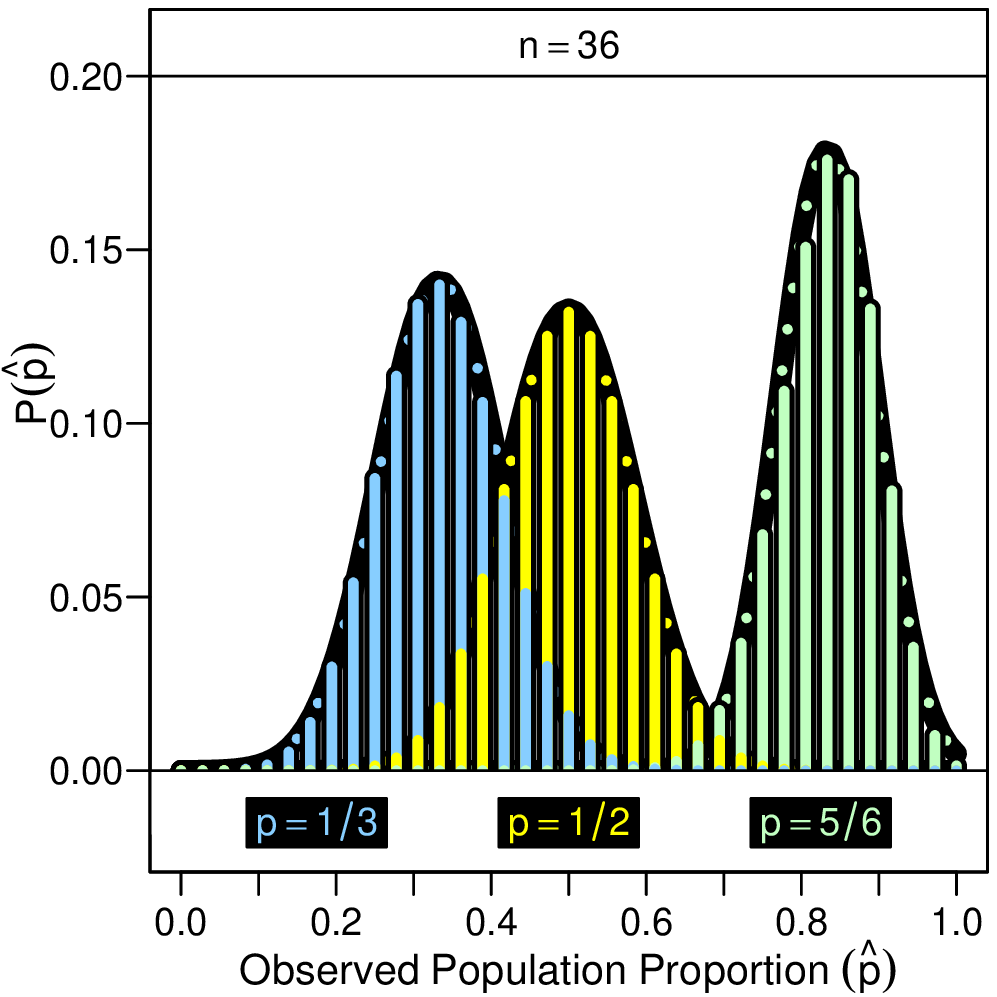}
\caption{Comparison between the true binomial distribution of the $\hat{p}$ statistic (i.e., the observed population proportion, $k/n$) and that assumed by the `normal approximation'.  Specifically, the $\hat{p}$ distributions of the binomial probability function with $p=1/3$, 1/2, and 5/6 are contrasted against scaled normal distributions with matching means and variances for sample sizes of $n=6$ (left panel) and $n=36$ (right panel), respectively.  In the small sample size regime the `normal approximation' provides a reasonable representation of the $\hat{p}$ distribution at $p=1/2$ and 1/3, but not 5/6, while in the intermediate-to-large sample size regime even the distribution at $p=5/6$ is also clearly converging towards normal.\label{fig3}}
\end{figure*}

\paragraph{A note on the prior}
The (non-informative) Bayes-Laplace uniform prior may in fact be viewed as the special case of $P_\mathrm{prior}(p)=B(1,1)=1$ within a wider family of possible conjugate priors for the binomial population proportion based on the beta distribution.  Another popular non-informative prior for $p$ is the Jeffreys' prior of $P_\mathrm{prior}(p)=B(1/2,1/2)$\footnote{Note that the factorial functions used in the beta distribution definition of Equation \ref{betaeqn} must be replaced by gamma functions according to the relation, $(m)!=\Gamma(m+1)$, in order to handle the non-integer input in this case.} (cf.\ \citealt{bro01,gel03}), which is, by design, proportional to the square root of the Fisher information.  Application of the Jeffreys' prior returns a posterior probability distribution for $p$ of $B(k+1/2,n-k+1/2)$.  The \textit{performance} of binomial CIs generated via beta distribution quantiles based on the Jeffreys' prior differ insignificantly from those based on the uniform prior over $0.025 \le p \le 0.975$ when $n \gtrsim 2$---consistent with the description of both these priors as `non-informative' (i.e., that even for small sample sizes the shape of the posterior probability distribution in both cases is strongly governed by the likelihood function of the observed data).  (See the recent review by \citealt{cou09} for a thorough evaluation of the performance of Bayesian CIs constructed with the Jeffreys' prior.) Hence, whilst the specific results presented in this paper are computed exclusively using the uniform prior, for the purposes of our general discussion regarding the superiority of the beta distribution quantile technique over the `normal approximation' and the \citet{clo34} approach these two non-informative priors may be considered interchangable. 

\section{The `Normal Approximation'}\label{section4}
For a system with an underlying binomial population proportion, $p$, neither very close to 0 or 1, one may suppose (with reference to the Central Limit Theorem) that the distribution of the $\hat{p}$ statistic in a series of independent samples of a fixed `large' size will follow approximately a normal distribution.  Under the assumptions of this `normal approximation' (also called the `Poisson error') one may employ the standard `Wald test' criterion, established by \citet{wal39}, to construct a two-sided confidence interval for $p$.  Specifically, at a confidence level of $c = 1-\alpha$ one may expect that the true value of $p$ lies within the interval:
\begin{equation}\label{gauss}
\hat{p}-z_{1-\alpha/2} \sqrt{\frac{\hat{p}\hat{q}}{n}} < p < \hat{p}+z_{1-\alpha/2} \sqrt{\frac{\hat{p}\hat{q}}{n}} 
\end{equation}
where $\hat{q}=1-\hat{p}$, and $z_{1-\alpha/2}$ is defined with reference to the standard normal distribution:
\[
\int_{-\infty}^{z_{1-\alpha/2}} \frac{1}{\sqrt{2 \pi}} \exp{(-x^2/2)} dx = 1-\alpha/2 \mathrm{\ .}
\]

Values of $z_{1-\alpha/2}$ for particular confidence levels may be obtained from reference tables in statistical textbooks (e.g.\ Quirin 1978) or computed within one's favourite mathematical software package (e.g.\ the \textsc{qnorm} function in \textsc{R}).  Of course, the most commonly used formula for constructing error bars on measured galaxy bar fractions, $p = \hat{p} \pm \sqrt{\hat{p}\hat{q}/n}$ (e.g.\ \citealt{elm90}), is simply the application of Equation \ref{gauss} at $z_{1-\alpha/2} = 1$, corresponding to a 1$\sigma$ confidence level of $c \approx 0.683$.  The cases of $z_{1-\alpha/2} = 2$ and 3 (i.e., 2$\sigma$ and 3$\sigma$ errors) correspond to higher confidence levels of $c \approx 0.954$ and 0.997, respectively.

As noted above, the key assumption behind this approach to binomial CI estimation---that the distribution of $\hat{p}$ may be approximated via a normal distribution with mean, $p$, and variance, $pq/n$---is reasonable only under the conditions of a `large' sample size and $p$ neither very close to 0 or 1.  In Figure \ref{fig3} I compare the distribution of the $\hat{p}$ statistic (computed directly from the binomial probability function) against the shape of the corresponding `normal approximation' for three different values of the underlying population proportion ($p = 1/3$, 1/2, and 5/6) and two different sample sizes ($n=6$ and 36).  In the small sample size example ($n=6$) the `normal approximation' provides a reasonable representation of the $\hat{p}$ distribution at $p=1/3$ and $p=1/2$, but performs poorly at $p=5/6$ (i.e., $p$ close to 1).  However, in the intermediate sample size example ($n=36$) there is now a clear convergence towards a normal distribution in $\hat{p}$ even at $p=5/6$.  These examples presented in Figure \ref{fig3} serve to illustrate the nature of deviations from `normality' in the distribution of $\hat{p}$ at small $n$ and/or extreme $p$ values.  The impact of these deviations on the performance of the `normal approximation' as a binomial CI generator is examined below.

\begin{figure*}
\includegraphics[width=15cm]{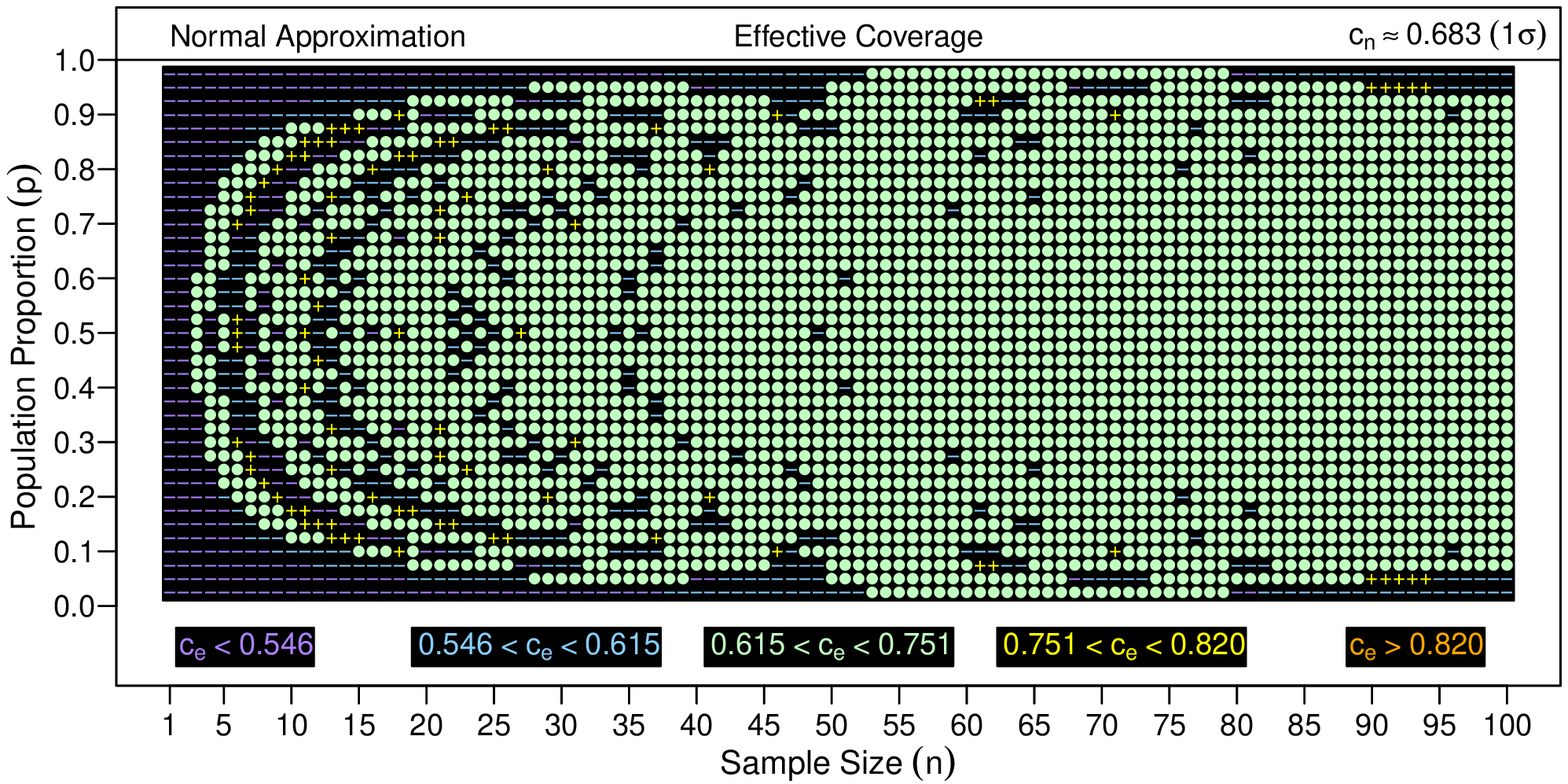} \includegraphics[width=15cm]{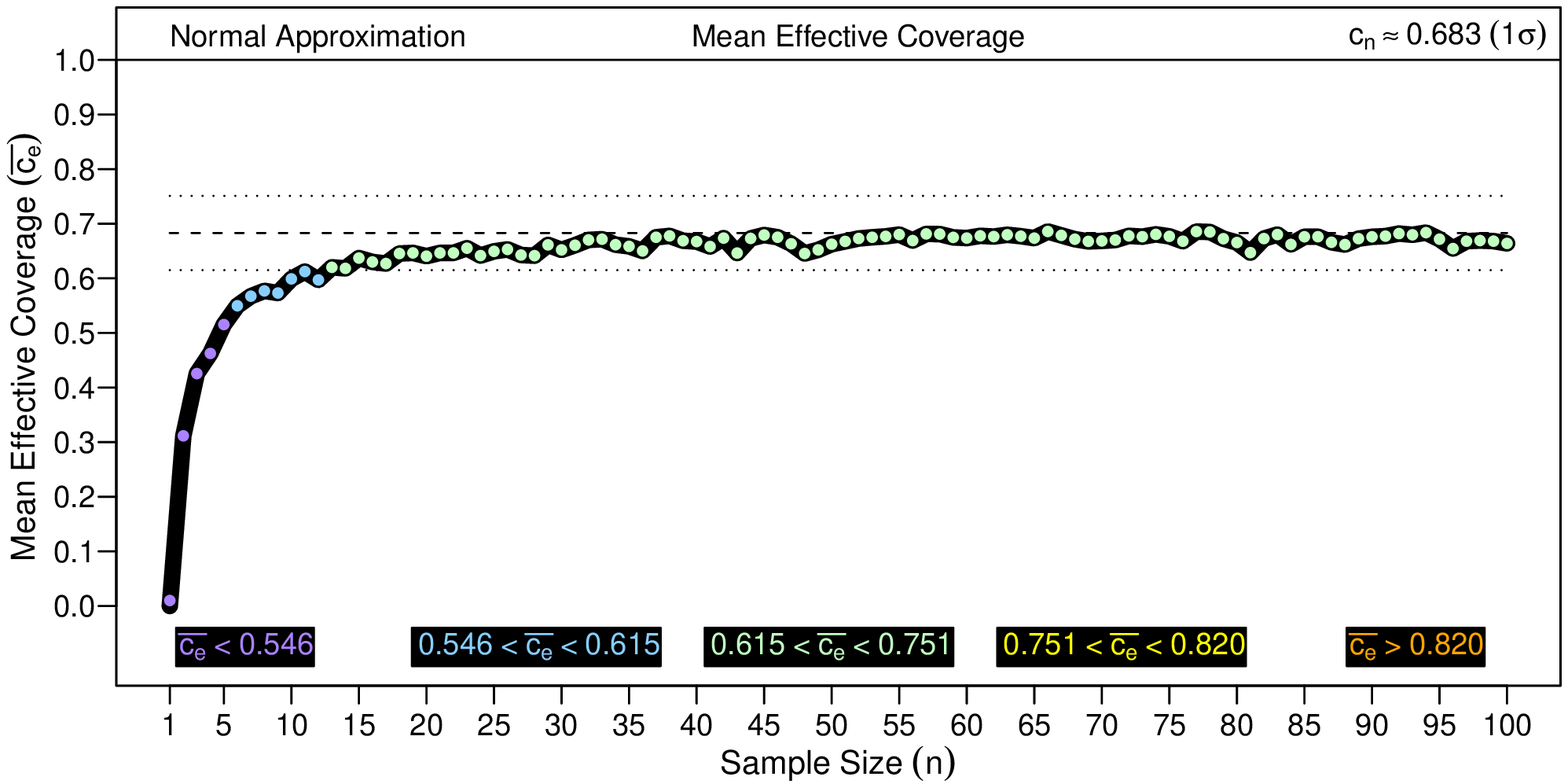}
\caption{The effective coverage, $c_e$, of confidence intervals on the binomial population proportion generated via the `normal approximation' at a nominal level of $c_n \approx 0.683$ (1$\sigma$) over the range $0.025 \le p \le 0.975$ and $1 \le n \le 100$ (upper panel).  Averaging the measured $c_e$ values unformly over all $p$ at each $n$ returns the \textit{mean} effective coverage as a function of sample size (lower panel).\label{fig4}}
\end{figure*}

In Figure \ref{fig4} I present the effective coverage of binomial CIs estimated via the `normal approximation' as a function of $p$ and $n$ at a nominal confidence level of $c_n \approx 0.683$ (1$\sigma$).  As in the case of the beta distribution quantile approach described above, there is a clear `oscillation signature' visible in this figure, reflecting a marked sensitivity in the coverage performance to the value of the underlying population proportion and sample size.\footnote{It is important also to note that this `oscillation signature' is evident even in binomial CIs generated via the `normal approximation' \textit{at very large sample sizes}, as thoroughly demonstrated by \citet{bro01,bro02}.  For instance, \citet{bro01} describe the erratic behaviour of the `normal approximation' coverage at a nominal level of $c_n = 0.95$ for a system with $p=0.005$, whereby there is a steady convergence in $c_e$ towards 0.95 for $n$ increasing until $n=592$, at which point coverage falls suddenly to $c_e=0.792$!  Similarly, \citet{bro02} demonstrate that in order to ensure coverage stays at or above a nominal level of $c_n=0.93$ for a system with $p=0.1$ using the `normal approximation' one requires a sample size of at least $n=286$, whereas for the Bayesian (Jeffreys' non-informative prior) CI this criterion is satisfied by $n=47$.}  However, it is also evident that the `normal approximation' suffers a \textit{systematic} decline in performance both for small $n$ and towards extreme values of $p$ near 0 or 1, generating binomial CIs with effective coverage far below the desired level.  The strict \textit{symmetry} of the `normal approximation' CI about the observed success fraction---which at low or high $\hat{p}$ may even extend (unphysically) to $p \le 0$ or $p \ge 1$---regardless of the inherent \textit{asymmetry} in the likelihood distribution for $p$ (see Figure \ref{fig1}) is the principal cause of these coverage failures.  The poor performance of the `normal approximation' at small $n$ is further highlighted in the corresponding plot of \textit{mean} effective coverage against sample size shown in the lower panel of Figure \ref{fig4}.  For the 1$\sigma$ CIs examined here (and popularly adopted in studies of the galaxy bar fraction) the \textit{mean} effective coverage of the `normal approximation' is far below the nominal level for $n \lesssim 20$, and should thus be strictly avoided in this small sample size regime.  Indeed, although its \textit{mean} effective coverage does ultimately improve with increasing $n$ one may be well advised to avoid the `normal approximation' altogether in light of its poor effective coverage at extreme $p$ values and the ready availability of a superior CI generator in the form of the (Bayesian) beta distribution quantiles described in Section \ref{section3}.

The flaws in the `normal approximation' as a CI generator described above were a great source of concern for statisticians in the 1930s, prompting the search for alternatives able to ensure universal coverage of at least the nominal level (thereby satisfying the classical definition of the term, `confidence interval'), whilst remaining readily computable given the limited aids available at the time (such as reference tables of quantiles for standard distributions).  The most popular of all such proposed alternatives was the \citet{clo34} approach (cf.\ \citealt{geh86}), which I review below.

\section{The Clopper \& Pearson Approach}\label{section5}
In their landmark 1934 paper Clopper \& Pearson presented a direct method for constructing `classical' confidence intervals on inferred population proportions based on quantiles of the binomial probability function (Equation \ref{binom}), guaranteed to provide a coverage probability of at least (but likely exceeding) the nominal confidence level.  The `two-sided' \citet{clo34} CI at $c=1-\alpha$ is constructed by solving the following equations for the lower and upper bounds, $P_l(k)=p_l$ and $P_u(k)=p_u$:
\begin{equation}\label{clo1}
\sum_{j=k}^{n} \binom{n}{j} p_l^{j} (1-p_l)^{n-j} = \alpha/2 \mathrm{\ (for\ }k\neq 0)
\end{equation}
and
\begin{equation}\label{clo2}
\sum_{j=0}^{k} \binom{n}{j} p_u^{j} (1-p_u)^{n-j} = \alpha/2 \mathrm{\ (for\ }k\neq n)
\end{equation}
where $k$ is again the observed number of successes (e.g.\ barred galaxies) in the sample, and $n$ the total sample size.  Note that in the extreme cases of $\hat{p}=0$ or 1, the \citet{clo34} formulae reduce simply to:
\begin{equation}\label{pzero}
p_l = (\alpha/2)^{1/n} \mathrm{\ for \ } \hat{p} = 1 \mathrm{\ and }
\end{equation}
\begin{equation}\label{pone}
p_u = 1-(\alpha/2)^{1/n} \mathrm{\ for \ } \hat{p} = 0 \mathrm{\ .}
\end{equation}

Modern mathematical software packages, such as \textsc{R} and \textsc{matlab}, support easy-to-use library functions (cf.\ \textsc{binom.test} in the \textsc{stats} package in \textsc{R}; or \textsc{binofit} in the \textsc{statistics toolbox} in \textsc{matlab}) for computation of \citet{clo34} confidence limits, which employ robust algorithms for solution of Equations \ref{clo1} and \ref{clo2}.  Alternatively, there exist numerous reference tables of pre-computed binomial CIs based on the \citet{clo34} approach---most notably those of \citet{geh86}, a popular reference for estimating uncertainties in astronomical population proportions.

\begin{figure*}
\includegraphics[width=15cm]{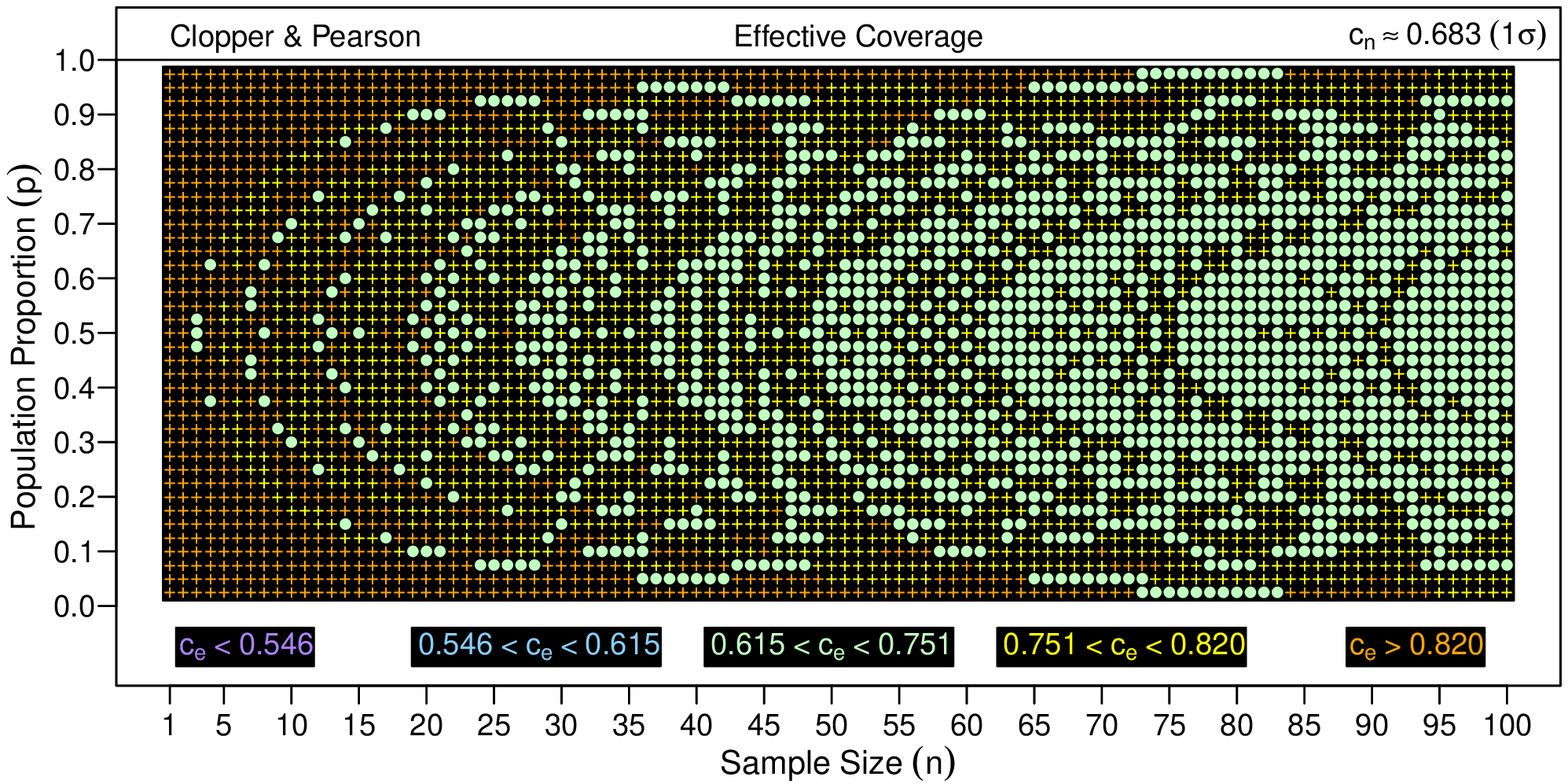} \includegraphics[width=15cm]{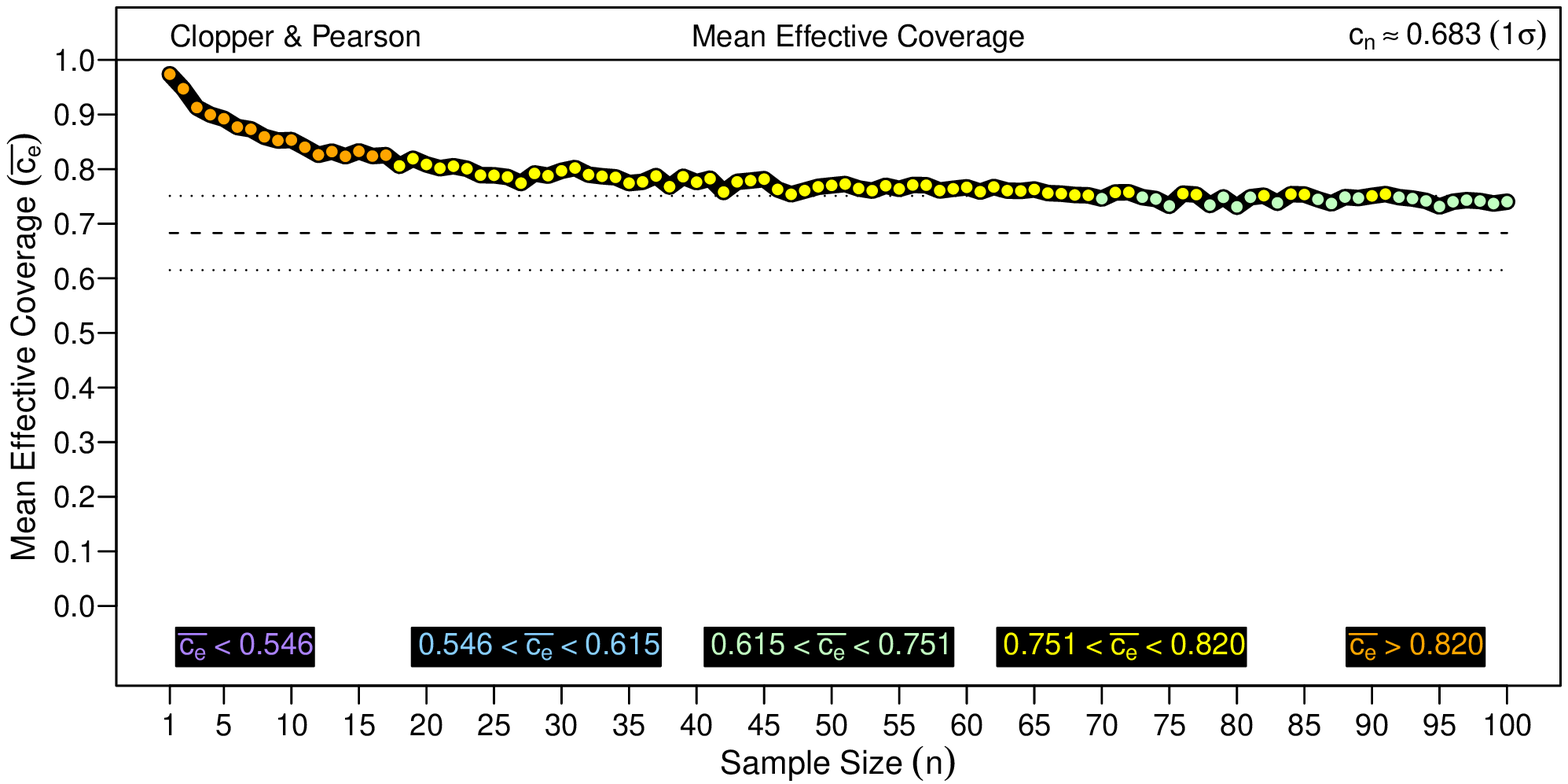}
\caption{The effective coverage, $c_e$, of confidence intervals on the binomial population proportion generated via the \citet{clo34} approach at a nominal level of $c_n \approx 0.683$ (1$\sigma$) over the range $0.025 \le p \le 0.975$ and $1 \le n \le 100$ (upper panel).  Averaging the measured $c_e$ values unformly over all $p$ at each $n$ returns the \textit{mean} effective coverage as a function of sample size (lower panel).\label{fig5}}
\end{figure*}

\begin{figure*}
\includegraphics[width=7.5cm]{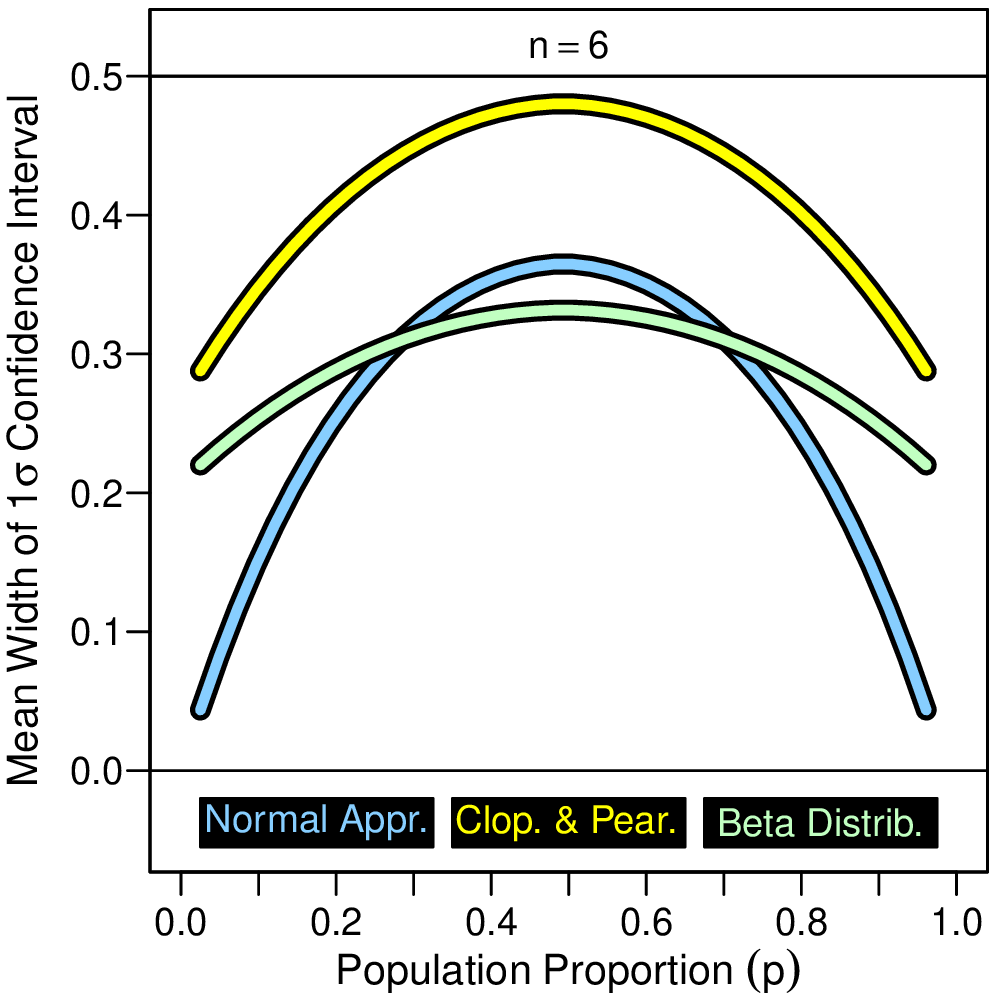} \includegraphics[width=7.5cm]{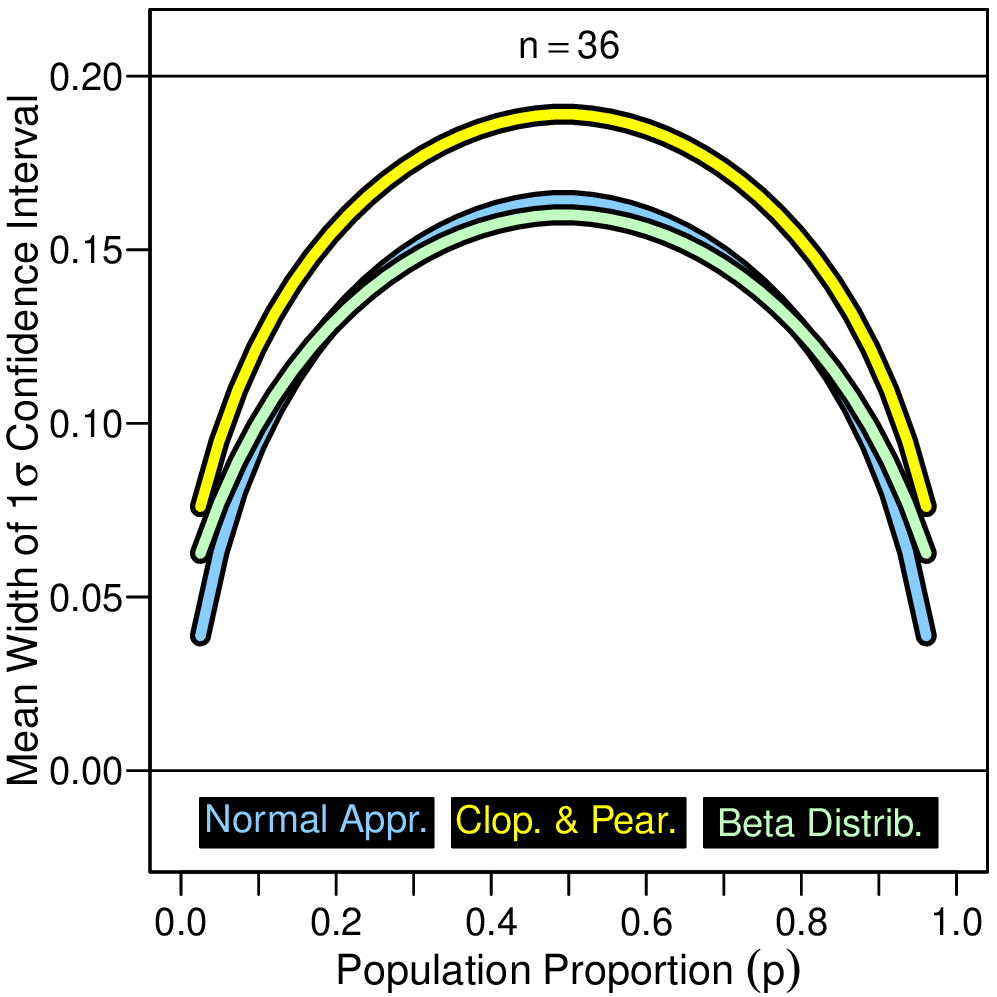}
\caption{Comparison between the mean widths of binomial CIs generated at $c \approx 0.683$ (1$\sigma$) via the beta distribution, the `normal approximation', and the \citet{clo34} approach, respectively, as a function of the underlying population proportion, $p$, for samples of sizes $n=6$ (left panel) and $n=36$ (right panel).\label{fig6}}
\end{figure*}

In the upper panel of Figure \ref{fig5} I examine the effective coverage of CIs generated via the \citet{clo34} approach as a function of $p$ and $n$ at a nominal confidence level of $c \approx 0.683$ (1$\sigma$).  In contrast with the results for both the beta distribution and the `normal approximation' reviewed above the \citet{clo34} CIs provide coverage far exceeding the nominal confidence level over much of this parameter space.  The \citet{clo34} coverage excess is also clearly evident in the corresponding \textit{mean} effective coverage for this CI generator plotted as a function of sample size in the lower panel of Figure \ref{fig5}.  Although the \citet{clo34} CIs do eventually converge to the nominal level at very large $n$, in the small-to-intermediate sample size regime their \textit{mean} effective coverage is consistently far above the desired level.  This point is in fact acknowledged in \citet{geh86}, although it appears not to be widely appreciated considering the frequency with which these CIs are treated as a `gold standard' in astronomical papers.

\section{Mean Confidence Interval Widths}
To illustrate the influence of the choice of CI generator on the estimated magnitude of the relevant uncertainties (i.e., the error bar size), I compare in Figure \ref{fig6} the \textit{mean} widths of $c\approx 0.683$ (1$\sigma$) CIs estimated via the (`equal-tailed') beta distribution quantile technique, the `normal approximation', and the \citet{clo34} approach as a function of $p$ for samples of sizes $n=6$ (left panel) and $n=36$ (right panel).  In the small sample size regime (where the `normal approximation' fails to provide sufficient coverage at $p \lesssim 1/6$ and $p \gtrsim 5/6$; see Figure \ref{fig4}) the mean CI widths are markedly smaller (by as much as $\Delta p \sim -0.15$) than those derived using the beta distribution technique (which generally provides superior coverage at these $p$ values; see Figure \ref{fig2}).  (Of course, the beta distribution should not be viewed as a strict benchmark for the ideal CI width, since its coverage is indeed prone to erratic performance at certain $p$ values---the `oscillation signature' to which \textit{all} non-randomising binomial CI generators are prone; although, as we have argued above, its performance may be considered the best of the three approaches examined in this study.)  In the intermediate sample size regime, the mean widths of these these two CI generators are in much better agreement, except at the extremes of $p \lesssim 1/20$ and $p \gtrsim 19/20$ where a marked under-estimation is still evident in the `normal approximation' CIs.  The \citet{clo34} CIs, on the other hand, exhibit a much greater mean width than those of the beta distribution or `normal approximation', regardless of $p$---reflecting the substantial coverage excess demonstrated for this CI generator in Section \ref{section5} (see Figure \ref{fig5}).   These examples verify that the choice of CI generator can indeed have a substantial impact on the magnitude of the estimated uncertainties, thereby confirming this choice to be an important practical consideration for effective astronomical data analysis.

\section{Conclusions}
I have reviewed the performance of three alternative methods for estimating confidence intervals on binomial population proportions; namely, the beta distribution quantile technique, the `normal approximation', and the \citet{clo34} approach (cf.\ \citealt{geh86}).  Despite their current popularity in astronomical research, the latter two CI generators are demonstrated to perform poorly under sampling conditions routinely encountered in observational studies---with the `normal approximation' failing to provide CIs of sufficient width to achieve coverage at the nominal confidence level, and the \citet{clo34} approach producing CIs far wider than necessary to achieve the nominal coverage.  In contrast, the (Bayesian) beta distribution quantile technique, is revealed to be a well-motivated alternative, consistently providing a mean level of coverage close to the nominal level, even for small-to-intermediate sample sizes.  Given that the beta distribution generator for binomial CIs may be easily implemented using modern mathematical software packages, I advocate strongly that this technique be adopted in future studies aiming to constrain the true values of astronomical propulation proportions (e.g.\ the galaxy bar fraction, red sequence fraction, or merger fraction).

\appendix
\section{CI Code Fragments \& CI Reference Tables}
Here I provide simple code fragments demonstrating the implementation of the beta distribution CI generator via standard library routines in \textsc{R}, \textsc{matlab}, \textsc{mathematica}, \textsc{IDL}, and \textsc{python}.  The correct performance of these code fragments in one's preferred mathematical software package may be verified by comparison against the reference tables of binomial CIs presented here in Tables \ref{tableA1} and \ref{tableA2}.  As in the main body of this paper I denote the nominal confidence level, $c$, the observed success count, $k$, and the sample size, $n$.  In the following it is assumed that these variables have already been defined computationally by the user with $c$ a real/double, and $k$ and $n$ integers.\\

In the \textsc{R} statistical package:\\
\texttt{p\_lower $<$- qbeta((1-c)/2,k+1,n-k+1)}\\
\texttt{p\_upper $<$- qbeta(1-(1-c)/2,k+1,n-k+1)}\\

In \textsc{matlab}:\\
\texttt{p\_lower = betaincinv((1-c)/2,k+1,n-k+1)}\\
\texttt{p\_upper = betaincinv(1-(1-c)/2,k+1,n-k+1)}\\

In \textsc{mathematica}:\\
\texttt{plower = Quantile[BetaDistribution[k+1,n-k+1],(1-c)/2]}\\
\texttt{pupper = Quantile[BetaDistribution[k+1,n-k+1],1-(1-c)/2]}\\

In \textsc{IDL} (if an `IDL Analyst' license is available):\\
\texttt{p\_lower = IMSL\_BETACDF((1-c)/2,k+1,n-k+1,/INVERSE)}\\
\texttt{p\_upper = IMSL\_BETACDF(1-(1-c)/2,k+1,n-k+1,/INVERSE)}\\
otherwise, iteratively:\\
\texttt{z = FINDGEN(10000)*0.0001}\\
\texttt{Beta = IBETA(k+1,n-k+1,z)}\\
\texttt{il = VALUE\_LOCATE(Beta,(1-c)/2)}\\
\texttt{iu = VALUE\_LOCATE(Beta,1-(1-c)/2)}\\
\texttt{p\_lower = z[il]}\\
\texttt{p\_upper = z[ul]}\\

In \textsc{python}:\\
\texttt{import scipy.stats.distributions as dist}\\
\texttt{p\_lower = dist.beta.ppf((1-c)/2.,k+1,n-k+1)}\\
\texttt{p\_upper = dist.beta.ppf(1-(1-c)/2.,k+1,n-k+1)}\\

\begin{sidewaystable*}
\caption{Confidence interval estimates at $c \approx 0.683$ (1$\sigma$) on binomial population proportions from quantiles of the beta distribution for all possible observed success counts for sample sizes up to 20\label{tableA1}}
\vspace{-0.25cm}
\begin{center}
\begin{tabular}{cccccccccccccccccccccc}
\hline\hline
$n$ & $k=$0 & 1 & 2 &3&4&5&6&7&8&9&10&11&12&13&14&15&16&17&18&19&20\\
\hline
1  & ${}_{0.083}^{0.602}$ & ${}_{0.398}^{0.917}$ & $\ldots$ & $\ldots$ & $\ldots$ & $\ldots$ & $\ldots$ & $\ldots$ & $\ldots$ & $\ldots$ & $\ldots$ & $\ldots$ & $\ldots$ & $\ldots$ & $\ldots$ & $\ldots$ & $\ldots$ & $\ldots$ & $\ldots$ & $\ldots$ & $\ldots$\\ 
2  & ${}_{0.056}^{0.459}$ & ${}_{0.252}^{0.748}$ & ${}_{0.541}^{0.944}$ & $\ldots$ & $\ldots$ & $\ldots$ & $\ldots$ & $\ldots$ & $\ldots$ & $\ldots$ & $\ldots$ & $\ldots$ & $\ldots$ & $\ldots$ & $\ldots$ & $\ldots$ & $\ldots$ & $\ldots$ & $\ldots$ & $\ldots$ & $\ldots$\\ 
3  & ${}_{0.042}^{0.369}$ & ${}_{0.185}^{0.618}$ & ${}_{0.382}^{0.815}$ & ${}_{0.631}^{0.958}$ & $\ldots$ & $\ldots$ & $\ldots$ & $\ldots$ & $\ldots$ & $\ldots$ & $\ldots$ & $\ldots$ & $\ldots$ & $\ldots$ & $\ldots$ & $\ldots$ & $\ldots$ & $\ldots$ & $\ldots$ & $\ldots$ & $\ldots$\\ 
4  & ${}_{0.034}^{0.308}$ & ${}_{0.147}^{0.524}$ & ${}_{0.297}^{0.703}$ & ${}_{0.476}^{0.853}$ & ${}_{0.692}^{0.966}$ & $\ldots$ & $\ldots$ & $\ldots$ & $\ldots$ & $\ldots$ & $\ldots$ & $\ldots$ & $\ldots$ & $\ldots$ & $\ldots$ & $\ldots$ & $\ldots$ & $\ldots$ & $\ldots$ & $\ldots$ & $\ldots$\\ 
5  & ${}_{0.028}^{0.264}$ & ${}_{0.121}^{0.454}$ & ${}_{0.243}^{0.615}$ & ${}_{0.385}^{0.757}$ & ${}_{0.546}^{0.879}$ & ${}_{0.736}^{0.972}$ & $\ldots$ & $\ldots$ & $\ldots$ & $\ldots$ & $\ldots$ & $\ldots$ & $\ldots$ & $\ldots$ & $\ldots$ & $\ldots$ & $\ldots$ & $\ldots$ & $\ldots$ & $\ldots$ & $\ldots$\\ 
6  & ${}_{0.024}^{0.231}$ & ${}_{0.104}^{0.400}$ & ${}_{0.206}^{0.546}$ & ${}_{0.324}^{0.676}$ & ${}_{0.454}^{0.794}$ & ${}_{0.600}^{0.896}$ & ${}_{0.769}^{0.976}$ & $\ldots$ & $\ldots$ & $\ldots$ & $\ldots$ & $\ldots$ & $\ldots$ & $\ldots$ & $\ldots$ & $\ldots$ & $\ldots$ & $\ldots$ & $\ldots$ & $\ldots$ & $\ldots$\\ 
7  & ${}_{0.021}^{0.206}$ & ${}_{0.090}^{0.357}$ & ${}_{0.179}^{0.490}$ & ${}_{0.280}^{0.610}$ & ${}_{0.390}^{0.720}$ & ${}_{0.510}^{0.821}$ & ${}_{0.643}^{0.910}$ & ${}_{0.794}^{0.979}$ & $\ldots$ & $\ldots$ & $\ldots$ & $\ldots$ & $\ldots$ & $\ldots$ & $\ldots$ & $\ldots$ & $\ldots$ & $\ldots$ & $\ldots$ & $\ldots$ & $\ldots$\\ 
8  & ${}_{0.019}^{0.185}$ & ${}_{0.080}^{0.323}$ & ${}_{0.158}^{0.444}$ & ${}_{0.246}^{0.555}$ & ${}_{0.342}^{0.658}$ & ${}_{0.445}^{0.754}$ & ${}_{0.556}^{0.842}$ & ${}_{0.677}^{0.920}$ & ${}_{0.815}^{0.981}$ & $\ldots$ & $\ldots$ & $\ldots$ & $\ldots$ & $\ldots$ & $\ldots$ & $\ldots$ & $\ldots$ & $\ldots$ & $\ldots$ & $\ldots$ & $\ldots$\\ 
9  & ${}_{0.017}^{0.168}$ & ${}_{0.072}^{0.294}$ & ${}_{0.142}^{0.405}$ & ${}_{0.220}^{0.508}$ & ${}_{0.305}^{0.605}$ & ${}_{0.395}^{0.695}$ & ${}_{0.492}^{0.780}$ & ${}_{0.595}^{0.858}$ & ${}_{0.706}^{0.928}$ & ${}_{0.832}^{0.983}$ & $\ldots$ & $\ldots$ & $\ldots$ & $\ldots$ & $\ldots$ & $\ldots$ & $\ldots$ & $\ldots$ & $\ldots$ & $\ldots$ & $\ldots$\\ 
10  & ${}_{0.016}^{0.154}$ & ${}_{0.065}^{0.270}$ & ${}_{0.128}^{0.373}$ & ${}_{0.199}^{0.469}$ & ${}_{0.275}^{0.559}$ & ${}_{0.356}^{0.644}$ & ${}_{0.441}^{0.725}$ & ${}_{0.531}^{0.801}$ & ${}_{0.627}^{0.872}$ & ${}_{0.730}^{0.935}$ & ${}_{0.846}^{0.984}$ & $\ldots$ & $\ldots$ & $\ldots$ & $\ldots$ & $\ldots$ & $\ldots$ & $\ldots$ & $\ldots$ & $\ldots$ & $\ldots$\\ 
11  & ${}_{0.014}^{0.142}$ & ${}_{0.060}^{0.250}$ & ${}_{0.117}^{0.346}$ & ${}_{0.181}^{0.435}$ & ${}_{0.250}^{0.519}$ & ${}_{0.324}^{0.600}$ & ${}_{0.400}^{0.676}$ & ${}_{0.481}^{0.750}$ & ${}_{0.565}^{0.819}$ & ${}_{0.654}^{0.883}$ & ${}_{0.750}^{0.940}$ & ${}_{0.858}^{0.986}$ & $\ldots$ & $\ldots$ & $\ldots$ & $\ldots$ & $\ldots$ & $\ldots$ & $\ldots$ & $\ldots$ & $\ldots$\\ 
12  & ${}_{0.013}^{0.132}$ & ${}_{0.055}^{0.232}$ & ${}_{0.108}^{0.322}$ & ${}_{0.167}^{0.405}$ & ${}_{0.230}^{0.485}$ & ${}_{0.297}^{0.561}$ & ${}_{0.366}^{0.634}$ & ${}_{0.439}^{0.703}$ & ${}_{0.515}^{0.770}$ & ${}_{0.595}^{0.833}$ & ${}_{0.678}^{0.892}$ & ${}_{0.768}^{0.945}$ & ${}_{0.868}^{0.987}$ & $\ldots$ & $\ldots$ & $\ldots$ & $\ldots$ & $\ldots$ & $\ldots$ & $\ldots$ & $\ldots$\\ 
13  & ${}_{0.012}^{0.123}$ & ${}_{0.051}^{0.217}$ & ${}_{0.100}^{0.301}$ & ${}_{0.154}^{0.380}$ & ${}_{0.213}^{0.455}$ & ${}_{0.274}^{0.526}$ & ${}_{0.338}^{0.595}$ & ${}_{0.405}^{0.662}$ & ${}_{0.474}^{0.726}$ & ${}_{0.545}^{0.787}$ & ${}_{0.620}^{0.846}$ & ${}_{0.699}^{0.900}$ & ${}_{0.783}^{0.949}$ & ${}_{0.877}^{0.988}$ & $\ldots$ & $\ldots$ & $\ldots$ & $\ldots$ & $\ldots$ & $\ldots$ & $\ldots$\\ 
14  & ${}_{0.011}^{0.116}$ & ${}_{0.048}^{0.204}$ & ${}_{0.093}^{0.283}$ & ${}_{0.144}^{0.357}$ & ${}_{0.198}^{0.428}$ & ${}_{0.255}^{0.496}$ & ${}_{0.314}^{0.561}$ & ${}_{0.375}^{0.625}$ & ${}_{0.439}^{0.686}$ & ${}_{0.504}^{0.745}$ & ${}_{0.572}^{0.802}$ & ${}_{0.643}^{0.856}$ & ${}_{0.717}^{0.907}$ & ${}_{0.796}^{0.952}$ & ${}_{0.884}^{0.989}$ & $\ldots$ & $\ldots$ & $\ldots$ & $\ldots$ & $\ldots$ & $\ldots$\\ 
15  & ${}_{0.011}^{0.109}$ & ${}_{0.045}^{0.192}$ & ${}_{0.087}^{0.267}$ & ${}_{0.134}^{0.337}$ & ${}_{0.185}^{0.404}$ & ${}_{0.238}^{0.469}$ & ${}_{0.293}^{0.531}$ & ${}_{0.350}^{0.592}$ & ${}_{0.408}^{0.650}$ & ${}_{0.469}^{0.707}$ & ${}_{0.531}^{0.762}$ & ${}_{0.596}^{0.815}$ & ${}_{0.663}^{0.866}$ & ${}_{0.733}^{0.913}$ & ${}_{0.808}^{0.955}$ & ${}_{0.891}^{0.989}$ & $\ldots$ & $\ldots$ & $\ldots$ & $\ldots$ & $\ldots$\\ 
16  & ${}_{0.010}^{0.103}$ & ${}_{0.042}^{0.181}$ & ${}_{0.082}^{0.252}$ & ${}_{0.126}^{0.319}$ & ${}_{0.174}^{0.383}$ & ${}_{0.223}^{0.444}$ & ${}_{0.274}^{0.504}$ & ${}_{0.327}^{0.562}$ & ${}_{0.382}^{0.618}$ & ${}_{0.438}^{0.673}$ & ${}_{0.496}^{0.726}$ & ${}_{0.556}^{0.777}$ & ${}_{0.617}^{0.826}$ & ${}_{0.681}^{0.874}$ & ${}_{0.748}^{0.918}$ & ${}_{0.819}^{0.958}$ & ${}_{0.897}^{0.990}$ & $\ldots$ & $\ldots$ & $\ldots$ & $\ldots$\\ 
17  & ${}_{0.010}^{0.097}$ & ${}_{0.040}^{0.172}$ & ${}_{0.077}^{0.239}$ & ${}_{0.119}^{0.303}$ & ${}_{0.164}^{0.363}$ & ${}_{0.210}^{0.422}$ & ${}_{0.258}^{0.479}$ & ${}_{0.308}^{0.534}$ & ${}_{0.359}^{0.588}$ & ${}_{0.412}^{0.641}$ & ${}_{0.466}^{0.692}$ & ${}_{0.521}^{0.742}$ & ${}_{0.578}^{0.790}$ & ${}_{0.637}^{0.836}$ & ${}_{0.697}^{0.881}$ & ${}_{0.761}^{0.923}$ & ${}_{0.828}^{0.960}$ & ${}_{0.903}^{0.990}$ & $\ldots$ & $\ldots$ & $\ldots$\\ 
18  & ${}_{0.009}^{0.092}$ & ${}_{0.038}^{0.163}$ & ${}_{0.073}^{0.228}$ & ${}_{0.113}^{0.288}$ & ${}_{0.155}^{0.346}$ & ${}_{0.198}^{0.402}$ & ${}_{0.244}^{0.456}$ & ${}_{0.291}^{0.509}$ & ${}_{0.339}^{0.561}$ & ${}_{0.388}^{0.612}$ & ${}_{0.439}^{0.661}$ & ${}_{0.491}^{0.709}$ & ${}_{0.544}^{0.756}$ & ${}_{0.598}^{0.802}$ & ${}_{0.654}^{0.845}$ & ${}_{0.712}^{0.887}$ & ${}_{0.772}^{0.927}$ & ${}_{0.837}^{0.962}$ & ${}_{0.908}^{0.991}$ & $\ldots$ & $\ldots$\\ 
19  & ${}_{0.009}^{0.088}$ & ${}_{0.036}^{0.156}$ & ${}_{0.070}^{0.217}$ & ${}_{0.107}^{0.275}$ & ${}_{0.147}^{0.330}$ & ${}_{0.188}^{0.384}$ & ${}_{0.231}^{0.436}$ & ${}_{0.275}^{0.487}$ & ${}_{0.321}^{0.537}$ & ${}_{0.367}^{0.585}$ & ${}_{0.415}^{0.633}$ & ${}_{0.463}^{0.679}$ & ${}_{0.513}^{0.725}$ & ${}_{0.564}^{0.769}$ & ${}_{0.616}^{0.812}$ & ${}_{0.670}^{0.853}$ & ${}_{0.725}^{0.893}$ & ${}_{0.783}^{0.930}$ & ${}_{0.844}^{0.964}$ & ${}_{0.912}^{0.991}$ & $\ldots$\\ 
20  & ${}_{0.008}^{0.084}$ & ${}_{0.034}^{0.149}$ & ${}_{0.066}^{0.207}$ & ${}_{0.102}^{0.263}$ & ${}_{0.139}^{0.316}$ & ${}_{0.179}^{0.367}$ & ${}_{0.220}^{0.417}$ & ${}_{0.261}^{0.466}$ & ${}_{0.304}^{0.514}$ & ${}_{0.348}^{0.561}$ & ${}_{0.393}^{0.607}$ & ${}_{0.439}^{0.652}$ & ${}_{0.486}^{0.696}$ & ${}_{0.534}^{0.739}$ & ${}_{0.583}^{0.780}$ & ${}_{0.633}^{0.821}$ & ${}_{0.684}^{0.861}$ & ${}_{0.737}^{0.898}$ & ${}_{0.793}^{0.934}$ & ${}_{0.851}^{0.966}$ & ${}_{0.916}^{0.992}$ \\ 
\hline
\end{tabular}
\end{center}
\end{sidewaystable*}

\begin{sidewaystable*}
\caption{Confidence interval estimates at $c \approx 0.997$ (3$\sigma$) on binomial population proportions from quantiles of the beta distribution for all possible observed success counts for sample sizes up to 20\label{tableA2}}
\vspace{-0.25cm}
\begin{center}
\begin{tabular}{cccccccccccccccccccccc}
\hline\hline
$n$ & $k=$0 & 1 & 2 &3&4&5&6&7&8&9&10&11&12&13&14&15&16&17&18&19&20\\
\hline
 1  & ${}_{0.001}^{0.963}$ & ${}_{0.037}^{0.999}$ & $\ldots$ & $\ldots$ & $\ldots$ & $\ldots$ & $\ldots$ & $\ldots$ & $\ldots$ & $\ldots$ & $\ldots$ & $\ldots$ & $\ldots$ & $\ldots$ & $\ldots$ & $\ldots$ & $\ldots$ & $\ldots$ & $\ldots$ & $\ldots$ & $\ldots$\\ 
 2  & ${}_{0.000}^{0.889}$ & ${}_{0.021}^{0.979}$ & ${}_{0.111}^{1.000}$ & $\ldots$ & $\ldots$ & $\ldots$ & $\ldots$ & $\ldots$ & $\ldots$ & $\ldots$ & $\ldots$ & $\ldots$ & $\ldots$ & $\ldots$ & $\ldots$ & $\ldots$ & $\ldots$ & $\ldots$ & $\ldots$ & $\ldots$ & $\ldots$\\ 
 3  & ${}_{0.000}^{0.808}$ & ${}_{0.015}^{0.929}$ & ${}_{0.071}^{0.985}$ & ${}_{0.192}^{1.000}$ & $\ldots$ & $\ldots$ & $\ldots$ & $\ldots$ & $\ldots$ & $\ldots$ & $\ldots$ & $\ldots$ & $\ldots$ & $\ldots$ & $\ldots$ & $\ldots$ & $\ldots$ & $\ldots$ & $\ldots$ & $\ldots$ & $\ldots$\\ 
 4  & ${}_{0.000}^{0.733}$ & ${}_{0.012}^{0.868}$ & ${}_{0.053}^{0.947}$ & ${}_{0.132}^{0.988}$ & ${}_{0.267}^{1.000}$ & $\ldots$ & $\ldots$ & $\ldots$ & $\ldots$ & $\ldots$ & $\ldots$ & $\ldots$ & $\ldots$ & $\ldots$ & $\ldots$ & $\ldots$ & $\ldots$ & $\ldots$ & $\ldots$ & $\ldots$ & $\ldots$\\ 
 5  & ${}_{0.000}^{0.668}$ & ${}_{0.010}^{0.807}$ & ${}_{0.042}^{0.898}$ & ${}_{0.102}^{0.958}$ & ${}_{0.193}^{0.990}$ & ${}_{0.332}^{1.000}$ & $\ldots$ & $\ldots$ & $\ldots$ & $\ldots$ & $\ldots$ & $\ldots$ & $\ldots$ & $\ldots$ & $\ldots$ & $\ldots$ & $\ldots$ & $\ldots$ & $\ldots$ & $\ldots$ & $\ldots$\\ 
 6  & ${}_{0.000}^{0.611}$ & ${}_{0.008}^{0.750}$ & ${}_{0.035}^{0.847}$ & ${}_{0.083}^{0.917}$ & ${}_{0.153}^{0.965}$ & ${}_{0.250}^{0.992}$ & ${}_{0.389}^{1.000}$ & $\ldots$ & $\ldots$ & $\ldots$ & $\ldots$ & $\ldots$ & $\ldots$ & $\ldots$ & $\ldots$ & $\ldots$ & $\ldots$ & $\ldots$ & $\ldots$ & $\ldots$ & $\ldots$\\ 
 7  & ${}_{0.000}^{0.562}$ & ${}_{0.007}^{0.698}$ & ${}_{0.030}^{0.797}$ & ${}_{0.070}^{0.872}$ & ${}_{0.128}^{0.930}$ & ${}_{0.203}^{0.970}$ & ${}_{0.302}^{0.993}$ & ${}_{0.438}^{1.000}$ & $\ldots$ & $\ldots$ & $\ldots$ & $\ldots$ & $\ldots$ & $\ldots$ & $\ldots$ & $\ldots$ & $\ldots$ & $\ldots$ & $\ldots$ & $\ldots$ & $\ldots$\\ 
 8  & ${}_{0.000}^{0.520}$ & ${}_{0.006}^{0.652}$ & ${}_{0.026}^{0.750}$ & ${}_{0.061}^{0.828}$ & ${}_{0.109}^{0.891}$ & ${}_{0.172}^{0.939}$ & ${}_{0.250}^{0.974}$ & ${}_{0.348}^{0.994}$ & ${}_{0.480}^{1.000}$ & $\ldots$ & $\ldots$ & $\ldots$ & $\ldots$ & $\ldots$ & $\ldots$ & $\ldots$ & $\ldots$ & $\ldots$ & $\ldots$ & $\ldots$ & $\ldots$\\ 
 9  & ${}_{0.000}^{0.484}$ & ${}_{0.006}^{0.610}$ & ${}_{0.023}^{0.707}$ & ${}_{0.054}^{0.785}$ & ${}_{0.096}^{0.851}$ & ${}_{0.149}^{0.904}$ & ${}_{0.215}^{0.946}$ & ${}_{0.293}^{0.977}$ & ${}_{0.390}^{0.994}$ & ${}_{0.516}^{1.000}$ & $\ldots$ & $\ldots$ & $\ldots$ & $\ldots$ & $\ldots$ & $\ldots$ & $\ldots$ & $\ldots$ & $\ldots$ & $\ldots$ & $\ldots$\\ 
 10  & ${}_{0.000}^{0.452}$ & ${}_{0.005}^{0.573}$ & ${}_{0.021}^{0.667}$ & ${}_{0.048}^{0.745}$ & ${}_{0.085}^{0.812}$ & ${}_{0.132}^{0.868}$ & ${}_{0.188}^{0.915}$ & ${}_{0.255}^{0.952}$ & ${}_{0.333}^{0.979}$ & ${}_{0.427}^{0.995}$ & ${}_{0.548}^{1.000}$ & $\ldots$ & $\ldots$ & $\ldots$ & $\ldots$ & $\ldots$ & $\ldots$ & $\ldots$ & $\ldots$ & $\ldots$ & $\ldots$\\ 
 11  & ${}_{0.000}^{0.423}$ & ${}_{0.005}^{0.540}$ & ${}_{0.019}^{0.632}$ & ${}_{0.044}^{0.708}$ & ${}_{0.077}^{0.775}$ & ${}_{0.118}^{0.832}$ & ${}_{0.168}^{0.882}$ & ${}_{0.225}^{0.923}$ & ${}_{0.292}^{0.956}$ & ${}_{0.368}^{0.981}$ & ${}_{0.460}^{0.995}$ & ${}_{0.577}^{1.000}$ & $\ldots$ & $\ldots$ & $\ldots$ & $\ldots$ & $\ldots$ & $\ldots$ & $\ldots$ & $\ldots$ & $\ldots$\\ 
 12  & ${}_{0.000}^{0.398}$ & ${}_{0.004}^{0.510}$ & ${}_{0.018}^{0.599}$ & ${}_{0.040}^{0.674}$ & ${}_{0.070}^{0.740}$ & ${}_{0.107}^{0.798}$ & ${}_{0.152}^{0.848}$ & ${}_{0.202}^{0.893}$ & ${}_{0.260}^{0.930}$ & ${}_{0.326}^{0.960}$ & ${}_{0.401}^{0.982}$ & ${}_{0.490}^{0.996}$ & ${}_{0.602}^{1.000}$ & $\ldots$ & $\ldots$ & $\ldots$ & $\ldots$ & $\ldots$ & $\ldots$ & $\ldots$ & $\ldots$\\ 
 13  & ${}_{0.000}^{0.376}$ & ${}_{0.004}^{0.484}$ & ${}_{0.016}^{0.569}$ & ${}_{0.037}^{0.643}$ & ${}_{0.064}^{0.707}$ & ${}_{0.098}^{0.765}$ & ${}_{0.138}^{0.816}$ & ${}_{0.184}^{0.862}$ & ${}_{0.235}^{0.902}$ & ${}_{0.293}^{0.936}$ & ${}_{0.357}^{0.963}$ & ${}_{0.431}^{0.984}$ & ${}_{0.516}^{0.996}$ & ${}_{0.624}^{1.000}$ & $\ldots$ & $\ldots$ & $\ldots$ & $\ldots$ & $\ldots$ & $\ldots$ & $\ldots$\\ 
 14  & ${}_{0.000}^{0.356}$ & ${}_{0.004}^{0.459}$ & ${}_{0.015}^{0.542}$ & ${}_{0.034}^{0.614}$ & ${}_{0.059}^{0.677}$ & ${}_{0.091}^{0.734}$ & ${}_{0.127}^{0.785}$ & ${}_{0.168}^{0.832}$ & ${}_{0.215}^{0.873}$ & ${}_{0.266}^{0.909}$ & ${}_{0.323}^{0.941}$ & ${}_{0.386}^{0.966}$ & ${}_{0.458}^{0.985}$ & ${}_{0.541}^{0.996}$ & ${}_{0.644}^{1.000}$ & $\ldots$ & $\ldots$ & $\ldots$ & $\ldots$ & $\ldots$ & $\ldots$\\ 
 15  & ${}_{0.000}^{0.338}$ & ${}_{0.003}^{0.438}$ & ${}_{0.014}^{0.517}$ & ${}_{0.032}^{0.587}$ & ${}_{0.055}^{0.649}$ & ${}_{0.084}^{0.705}$ & ${}_{0.118}^{0.756}$ & ${}_{0.155}^{0.802}$ & ${}_{0.198}^{0.845}$ & ${}_{0.244}^{0.882}$ & ${}_{0.295}^{0.916}$ & ${}_{0.351}^{0.945}$ & ${}_{0.413}^{0.968}$ & ${}_{0.483}^{0.986}$ & ${}_{0.562}^{0.997}$ & ${}_{0.662}^{1.000}$ & $\ldots$ & $\ldots$ & $\ldots$ & $\ldots$ & $\ldots$\\ 
 16  & ${}_{0.000}^{0.322}$ & ${}_{0.003}^{0.417}$ & ${}_{0.013}^{0.495}$ & ${}_{0.030}^{0.562}$ & ${}_{0.052}^{0.623}$ & ${}_{0.078}^{0.678}$ & ${}_{0.109}^{0.728}$ & ${}_{0.144}^{0.774}$ & ${}_{0.183}^{0.817}$ & ${}_{0.226}^{0.856}$ & ${}_{0.272}^{0.891}$ & ${}_{0.322}^{0.922}$ & ${}_{0.377}^{0.948}$ & ${}_{0.438}^{0.970}$ & ${}_{0.505}^{0.987}$ & ${}_{0.583}^{0.997}$ & ${}_{0.678}^{1.000}$ & $\ldots$ & $\ldots$ & $\ldots$ & $\ldots$\\ 
 17  & ${}_{0.000}^{0.307}$ & ${}_{0.003}^{0.399}$ & ${}_{0.012}^{0.474}$ & ${}_{0.028}^{0.539}$ & ${}_{0.048}^{0.598}$ & ${}_{0.074}^{0.652}$ & ${}_{0.102}^{0.702}$ & ${}_{0.135}^{0.748}$ & ${}_{0.171}^{0.790}$ & ${}_{0.210}^{0.829}$ & ${}_{0.252}^{0.865}$ & ${}_{0.298}^{0.898}$ & ${}_{0.348}^{0.926}$ & ${}_{0.402}^{0.952}$ & ${}_{0.461}^{0.972}$ & ${}_{0.526}^{0.988}$ & ${}_{0.601}^{0.997}$ & ${}_{0.693}^{1.000}$ & $\ldots$ & $\ldots$ & $\ldots$\\ 
 18  & ${}_{0.000}^{0.294}$ & ${}_{0.003}^{0.382}$ & ${}_{0.012}^{0.455}$ & ${}_{0.026}^{0.518}$ & ${}_{0.046}^{0.575}$ & ${}_{0.069}^{0.628}$ & ${}_{0.096}^{0.677}$ & ${}_{0.126}^{0.722}$ & ${}_{0.160}^{0.765}$ & ${}_{0.196}^{0.804}$ & ${}_{0.235}^{0.840}$ & ${}_{0.278}^{0.874}$ & ${}_{0.323}^{0.904}$ & ${}_{0.372}^{0.931}$ & ${}_{0.425}^{0.954}$ & ${}_{0.482}^{0.974}$ & ${}_{0.545}^{0.988}$ & ${}_{0.618}^{0.997}$ & ${}_{0.706}^{1.000}$ & $\ldots$ & $\ldots$\\ 
 19  & ${}_{0.000}^{0.281}$ & ${}_{0.003}^{0.367}$ & ${}_{0.011}^{0.437}$ & ${}_{0.025}^{0.498}$ & ${}_{0.043}^{0.554}$ & ${}_{0.065}^{0.606}$ & ${}_{0.091}^{0.654}$ & ${}_{0.119}^{0.698}$ & ${}_{0.150}^{0.740}$ & ${}_{0.184}^{0.779}$ & ${}_{0.221}^{0.816}$ & ${}_{0.260}^{0.850}$ & ${}_{0.302}^{0.881}$ & ${}_{0.346}^{0.909}$ & ${}_{0.394}^{0.935}$ & ${}_{0.446}^{0.957}$ & ${}_{0.502}^{0.975}$ & ${}_{0.563}^{0.989}$ & ${}_{0.633}^{0.997}$ & ${}_{0.719}^{1.000}$ & $\ldots$\\ 
 20  & ${}_{0.000}^{0.270}$ & ${}_{0.003}^{0.353}$ & ${}_{0.011}^{0.420}$ & ${}_{0.024}^{0.480}$ & ${}_{0.041}^{0.535}$ & ${}_{0.062}^{0.585}$ & ${}_{0.086}^{0.632}$ & ${}_{0.113}^{0.676}$ & ${}_{0.142}^{0.717}$ & ${}_{0.174}^{0.756}$ & ${}_{0.208}^{0.792}$ & ${}_{0.244}^{0.826}$ & ${}_{0.283}^{0.858}$ & ${}_{0.324}^{0.887}$ & ${}_{0.368}^{0.914}$ & ${}_{0.415}^{0.938}$ & ${}_{0.465}^{0.959}$ & ${}_{0.520}^{0.976}$ & ${}_{0.580}^{0.989}$ & ${}_{0.647}^{0.997}$ & ${}_{0.730}^{1.000}$ \\ 
\hline
\end{tabular}
\end{center}
\end{sidewaystable*}

\section*{Acknowledgments}
The author would like to thank Matthew Prescott for his assistance in defining the \textsc{IDL} code fragment, Carlos L\'opez for supplying the \textsc{python} code fragment, Roban Kramer for numerous helpful discussions on the role of statistics in astronomy, and the anonymous referee for a thorough reading of the paper and many insightful comments.


\end{document}